\pdfoutput=1
\documentclass[twocolumn,twocolappendix]{aastex701}

\usepackage{amsmath,amstext}
\usepackage{natbib}
\usepackage{graphicx}
\usepackage{booktabs}
\usepackage{xspace}
\newcommand{\grizli}{\textsc{Gri$z$li}\xspace}
\newcommand{\eazy}{\textsc{EA$z$Y}\xspace}
\newcommand{\pros}{\textsc{Prospector}\xspace}
\newcommand{\dynesty}{\textsc{dynesty}\xspace}
\newcommand{\galfit}{\textsc{GALFIT}}
\newcommand{\pysersic}{\textsc{pysersic}\xspace}
\newcommand{\jwst}{\textsc{jwst}\xspace}
\providecommand{\farcs}{\mbox{$.\!\!^{\prime\prime}$}}
\graphicspath{{figures/}}

\shorttitle{A sub-100~pc view of a multiply imaged quiescent galaxy at $z\sim5$}
\shortauthors{Pan et al.}

\begin{document}

\title{A sub-100~pc view at $z\sim5$ of a multiply imaged massive quiescent galaxy}

\correspondingauthor{Richard Pan}
\email{richard.pan@tufts.edu}

\author[0000-0002-9651-5716]{Richard Pan}
\affiliation{Department of Physics \& Astronomy, Tufts University, Medford, MA, USA}
\email[show]{richard.pan@tufts.edu}

\author[0000-0001-9002-3502]{Danilo Marchesini}
\affiliation{Department of Physics \& Astronomy, Tufts University, Medford, MA, USA}
\email{danilo.marchesini@tufts.edu}

\author[0000-0003-2718-8640]{Joseph F. V. Allingham}
\affiliation{Department of Physics, Ben-Gurion University of the Negev, Be'er-Sheva, Israel}
\email{allingha@post.bgu.ac.il}

\author[0000-0001-7201-5066]{Seiji Fujimoto}
\affiliation{David A. Dunlap Department of Astronomy and Astrophysics, University of Toronto, Toronto, Ontario, Canada}
\affiliation{Dunlap Institute for Astronomy and Astrophysics, Toronto, Ontario, Canada}
\email{seiji.fujimoto@utoronto.ca}

\author[0000-0002-0350-4488]{Adi Zitrin}
\affiliation{Department of Physics, Ben-Gurion University of the Negev, Be'er-Sheva, Israel}
\email{zitrin@bgu.ac.il}

\author[0000-0001-7410-7669]{Dan Coe}
\affiliation{Space Telescope Science Institute, Baltimore, MD, USA}
\affiliation{Association of Universities for Research in Astronomy (AURA), Inc. for the European Space Agency (ESA)}
\affiliation{Center for Astrophysical Sciences, Department of Physics and Astronomy, The Johns Hopkins University, Baltimore, MD, USA}
\email{dcoe@stsci.edu}

\author[0000-0002-5588-9156]{Vasily Kokorev}
\thanks{NHFP Hubble Fellow}
\affiliation{Department of Astronomy, The University of Texas at Austin, Austin, TX 78712, USA}
\affiliation{Cosmic Frontier Center, The University of Texas at Austin, Austin, TX 78712, USA}
\email{vasily.kokorev.astro@gmail.com}

\author[0000-0001-6278-032X]{Lukas J. Furtak}
\affiliation{Department of Astronomy, The University of Texas at Austin, Austin, TX, USA}
\affiliation{Cosmic Frontier Center, The University of Texas at Austin, Austin, TX, USA}
\email{furtak@utexas.edu}

\author{Jacqueline Antwi-Danso}
\affiliation{David A. Dunlap Department of Astronomy and Astrophysics, University of Toronto, Toronto, Ontario, Canada}
\affiliation{Dunlap Institute for Astronomy and Astrophysics, Toronto, Ontario, Canada}
\email{}

\author[0000-0003-3983-5438]{Yoshihisa Asada}
\affiliation{David A. Dunlap Department of Astronomy and Astrophysics, University of Toronto, Toronto, Ontario, Canada}
\affiliation{Dunlap Institute for Astronomy and Astrophysics, Toronto, Ontario, Canada}
\email{yoshi.asada@utoronto.ca}

\author[0000-0002-8686-8737]{Franz E. Bauer}
\affiliation{Instituto de Alta Investigaci\'on, Universidad de Tarapac\'a, Arica, Chile}
\email{fbauer@academicos.uta.cl}

\author[0000-0001-5984-0395]{Maru\v{s}a Brada\v{c}}
\affiliation{Faculty of Mathematics and Physics, University of Ljubljana, Ljubljana, Slovenia}
\affiliation{Department of Physics and Astronomy, University of California Davis, Davis, CA, USA}
\email{marusa.bradac@fmf.uni-lj.si}

\author[0000-0002-7908-9284]{Larry D. Bradley}
\affiliation{Space Telescope Science Institute, Baltimore, MD, USA}
\email{lbradley@stsci.edu}

\author{Chloe M. Cheng}
\affiliation{David A. Dunlap Department of Astronomy and Astrophysics, University of Toronto, Toronto, Ontario, Canada}
\affiliation{Dunlap Institute for Astronomy and Astrophysics, Toronto, Ontario, Canada}
\email{}

\author[0000-0001-8460-1564]{Pratika Dayal}
\affiliation{David A. Dunlap Department of Astronomy and Astrophysics, University of Toronto, Toronto, Ontario, Canada}
\affiliation{Dunlap Institute for Astronomy and Astrophysics, Toronto, Ontario, Canada}
\email{pratika.dayal@utoronto.ca}

\author[0000-0002-9382-9832]{Andreas L. Faisst}
\affiliation{IPAC, California Institute of Technology, Pasadena, CA, USA}
\email{afaisst@caltech.edu}

\author{Brenda L. Frye}
\affiliation{Department of Astronomy/Steward Observatory, University of Arizona, Tucson, AZ, USA}
\email{bfrye@arizona.edu}

\author[0000-0001-9440-8872]{Norman A. Grogin}
\affiliation{Space Telescope Science Institute, Baltimore, MD, USA}
\email{nagrogin@stsci.edu}

\author[0000-0003-4512-8705]{Tiger Yu-Yang Hsiao}
\affiliation{Department of Astronomy, The University of Texas at Austin, Austin, TX, USA}
\affiliation{Cosmic Frontier Center, The University of Texas at Austin, Austin, TX, USA}
\email{tiger.hsiao@utexas.edu}

\author[0000-0002-3475-7648]{Gourav Khullar}
\affiliation{Department of Astronomy, University of Washington, Seattle, WA, USA}
\affiliation{eScience Institute, University of Washington, Seattle, WA, USA}
\email{gkhullar@uw.edu}

\author[0000-0002-6610-2048]{Anton M. Koekemoer}
\affiliation{Space Telescope Science Institute, Baltimore, MD, USA}
\email{koekemoer@stsci.edu}

\author[0000-0003-2037-4619]{Conor Larison}
\affiliation{Space Telescope Science Institute, Baltimore, MD, USA}
\email{conorjlarison@gmail.com}

\author[0000-0003-1581-7825]{Ray A. Lucas}
\affiliation{Space Telescope Science Institute, Baltimore, MD, USA}
\email{lucas@stsci.edu}

\author[0000-0002-5694-6124]{Vladan Markov}
\affiliation{Faculty of Mathematics and Physics, University of Ljubljana, Ljubljana, Slovenia}
\email{vladan.markov@fmf.uni-lj.si}

\author[0000-0003-3243-9969]{Nicholas S. Martis}
\affiliation{Faculty of Mathematics and Physics, University of Ljubljana, Ljubljana, Slovenia}
\email{nicholas.martis@fmf.uni-lj.si}

\author[0009-0000-1999-5472]{Minami Nakane}
\affiliation{Institute for Cosmic Ray Research, The University of Tokyo, Kashiwa, Chiba, Japan}
\affiliation{Department of Physics, Graduate School of Science, The University of Tokyo, Tokyo, Japan}
\email{nakanem@icrr.u-tokyo.ac.jp}

\author[]{Ga\"el Noirot}
\affiliation{Space Telescope Science Institute, 3700 San Martin Drive, Baltimore, Maryland 21218, USA.}
\email{gnoirot@stsci.edu}

\author[0000-0001-5492-1049]{Johan Richard}
\affiliation{Universit\'e Claude Bernard Lyon 1, CRAL UMR5574, ENS de Lyon, CNRS, Villeurbanne, France}
\email{johan.richard@univ-lyon1.fr}

\author[0000-0003-4223-7324]{Massimo Ricotti}
\affiliation{Department of Astronomy, University of Maryland, College Park, MD, USA}
\email{ricotti@umd.edu}

\author[0000-0002-6265-2675]{Luke Robbins}
\affiliation{Department of Physics \& Astronomy, Tufts University, Medford, MA, USA}
\email{andrew.robbins@tufts.edu}

\author[0000-0002-4622-6617]{Fengwu Sun}
\affiliation{Center for Astrophysics $|$ Harvard \& Smithsonian, Cambridge, MA, USA}
\email{sunfengwu@westlake.edu.cn}

\author[0000-0001-6477-4011]{Francesco Valentino}
\affiliation{Cosmic Dawn Center (DAWN), Denmark}
\affiliation{DTU-Space, Technical University of Denmark, Kgs. Lyngby, Denmark}
\email{fmava@dtu.dk}

\author[0000-0002-5057-135X]{Eros Vanzella}
\affiliation{INAF -- OAS, Osservatorio di Astrofisica e Scienza dello Spazio di Bologna, Bologna, Italy}
\email{eros.vanzella@inaf.it}

\author[0000-0002-4201-7367]{Chris Willott}
\affiliation{NRC Herzberg, Victoria, BC, Canada}
\email{chris.willott@nrc.ca}

\author[0000-0001-8156-6281]{Rogier A. Windhorst}
\affiliation{School of Earth and Space Exploration, Arizona State University, Tempe, AZ, USA}
\email{Rogier.Windhorst@gmail.com}

\author[0009-0006-6763-4245]{Hiroto Yanagisawa}
\affiliation{Institute for Cosmic Ray Research, The University of Tokyo, Kashiwa, Chiba, Japan}
\affiliation{Department of Physics, Graduate School of Science, The University of Tokyo, Tokyo, Japan}
\email{yana@icrr.u-tokyo.ac.jp}

\author{Anita Zanella}
\affiliation{INAF -- OAS, Osservatorio di Astrofisica e Scienza dello Spazio di Bologna, Bologna, Italy}
\email{}

\begin{abstract}
Recent \jwst surveys have revealed massive quiescent galaxies that assembled rapidly and stopped forming stars, or quenched, within the first billion years. Simulations struggle to reproduce how quickly these galaxies formed and quenched, and models typically invoke a compact central starburst followed by feedback from a supermassive black hole. Observationally testing these models requires resolving the inner kiloparsec where this feedback should leave an imprint, but these galaxies are so compact that they appear unresolved, which biases their sizes, central densities, and ages. We present VENUS-CLJ0152-QG1, a massive post-starburst galaxy at $z\approx5.18$ that a foreground cluster lenses into three images, resolving it on 70 parsec scales. We find a half-light radius twice as large as expected and a central stellar mass density ten times lower than field analogs. These features suggest that it quenched without compaction or a central active galactic nucleus (AGN). Yet, we also find ionized gas emission offset from the stellar core, suggesting AGN outflows, shocks, or a recent merger. VCQG1 presents a puzzling scenario in which the first massive galaxies may quench without a primary compaction event. Ultimately, VCQG1 shows that lensing resolves the imprints of quenching (an extended, low-density structure, a younger core, and offset ionized gas) that would have blurred together in an unlensed field. Only with future resolved spectroscopy of such lensed galaxies can we confirm how the first massive galaxies formed and quenched, and which physical mechanisms drove these processes.
\end{abstract}


\submitjournal{ApJL}

\section{Introduction} \label{sec:intro}

Massive galaxies ($\log(M_*/M_\odot)\gtrsim10.5$) that have stopped forming stars, known as massive quiescent galaxies (MQGs), dominate the high-mass end of the galaxy population today \citep{pengMassEnvironmentDrivers2010, somervillePhysicalModelsGalaxy2015}. In the standard picture, they formed most of their stars early ($z>2$), quenched, and then grew in size through minor mergers that added little stellar mass \citep{naab_minor_2009, vandokkumGrowthMassiveGalaxies2010}. Their quenching is thought to be driven mainly by internal mechanisms, such as stellar winds, supernovae, and feedback from active galactic nuclei (AGN) that heats or expels the cold gas \citep{crotonManyLivesActive2006, hopkinsCosmologicalFrameworkCoEvolution2008, fabianObservationalEvidenceActive2012}. Studies at $z\lesssim3$ established this sequence by showing that many massive galaxies at $z\sim2$ were already quiescent but several times more compact than their local descendants \citep{vandokkumGrowthMassiveGalaxies2010, vandokkumFormingCompactMassive2015}. Because MQGs hold the stellar mass of the most massive galaxies and formed it first, when and how they formed and quenched sets how efficiently the earliest galaxies converted their gas into stars.

The \textit{James Webb Space Telescope} (\jwst) has pushed this picture to much earlier times. It now confirms MQGs spectroscopically out to $z\sim5$ and beyond \citep{carnall_massive_2023, glazebrook_massive_2024, nanayakkara_population_2024, carnall_jwst_2024, ito_deepdive_2025, de_graaff_efficient_2025, weibel_rubies_2025}, in larger numbers than most cosmological simulations predict \citep{carnallSurprisingAbundanceMassive2023, lagos_quenching_2024, baker_abundance_2025, antwi-danso_feniks_2025, weibel_rubies_2025}, and their stellar populations imply that many formed most of their mass within the first few hundred million years \citep{glazebrook_massive_2024, carnall_jwst_2024, de_graaff_efficient_2025, weibel_rubies_2025}. To reproduce them, simulations need either an extremely efficient conversion of baryons into stars \citep{glazebrook_massive_2024, de_graaff_efficient_2025} or volumes large enough to contain such rare objects \citep{baker_abundance_2025, chandro2025}. Either the physics in current models is incomplete, or most simulated volumes are too small to capture the observed abundance. Within these models, massive galaxies typically quench through a sequence in which gas-rich mergers or inflows drive gas to the center (compaction), fuel a central starburst, and grow a supermassive black hole whose feedback then removes or heats the remaining gas \citep{hopkinsRelationQuasarMerging2006, zolotovCompactionQuenchingHighz2015, kimmig_blowing_2025, ni_astrid_2025, saputraharyanaInterplayCompactionQuenching2026, chandro-gomez_unveiling_2026}.

Testing this sequence requires resolving the region where the feedback imprint is most noticeable, the inner kiloparsec. At $z<3$, quiescence is tightly linked to the stellar mass surface density within the central kiloparsec \citep{barro_structural_2017, whitaker_predicting_2017}, and radial gradients in the stellar populations separate inside-out, outside-in, and uniform quenching \citep{suess_color_2020, edward_magaz3ne_2026, whitaker_quenching_2026}. The MQGs found so far at $z>4$ appear extremely compact, with high central densities and, in several cases, signatures of a central starburst or AGN \citep{carnall_massive_2023, ji_reconstructing_2023, de_graaff_efficient_2025, weibel_rubies_2025}. Yet their sizes ($\sim0.2-0.6$\,kpc) are comparable to the resolution of NIRCam at the wavelengths that trace their stellar mass \citep{kawinwanichakij_connecting_2026}, and unresolved light profiles can overestimate their central densities, total stellar masses, and ages \citep{cutler_unbreaking_2026}. Current observations therefore cannot test the central prediction of these models. Did the first MQGs quench through compaction and central feedback, and what imprint did quenching leave on their inner kiloparsec?

Strong gravitational lensing offers a way to resolve these cores, by magnifying and stretching background galaxies. Resolved studies of lensed quiescent galaxies at Cosmic Noon have measured their rotation, chemical abundances, radial gradients, and resolved star-formation histories \citep{newman_discovery_2015, newmanResolvingQuiescentGalaxies2018a, jafariyazani_resolved_2020, akhshik_requiem-2d_2023, jafariyazaniChemicalAbundancesEarly2025}, and such analyses are now reaching earlier epochs \citep{khullarCOOLLAMPSExtraordinarilyBright2021, hutchisonJWSTWazArc2025}. Highly magnified MQGs are rare, so finding them at $z>4$ requires imaging many lensing clusters to probe larger volumes. The \jwst\ Vast Exploration for Nascent, Unexplored Sources (VENUS) Survey (GO-6882) images more than 60 lensing clusters in up to 10 NIRCam bands and covers a large volume of highly magnified regions. With VENUS, we can find MQGs at $z>4$ magnified enough to look inside their inner kiloparsec, map their stellar populations, and understand their potential evolutionary pathways.

We report the discovery of VENUS-CLJ0152-QG1 (hereafter VCQG1), a massive ($M_{*} = 3.36^{+0.64}_{-0.47} \times 10^{10}\,M_{\odot}$) galaxy strongly lensed by the foreground galaxy cluster CL J0152.7-1357 into three distinct images, identified in VENUS. The lensing magnification gives a physical resolution of about 70\,pc, which lets us measure the structure, colors, and ionized gas of a $z\sim5$ MQG on sub-kiloparsec scales. We describe the data and methods in Section~\ref{sec:methods}, present our measurements in Section~\ref{sec:results}, and place them in the context of other observations and simulations in Section~\ref{sec:discussion}, which ends with the caveats and future work (Section~\ref{subsec:caveats}). All magnitudes are given in the AB system \citep{oke_secondary_1983} and we assume a WMAP9 cosmology \citep{hinshaw_nine-year_2013} with $H_0=69.32~{\rm km~s^{-1}~Mpc^{-1}}$, $\Omega_M=0.2865$, and $\Omega_\Lambda=0.7134$, along with a \citet{chabrier_galactic_2003} initial mass function.

\section{Data and Methods} \label{sec:methods}

\subsection{Photometric data and Voronoi binning}
\label{subsec:data}

\begin{figure*}[t!]
	\centering
	\includegraphics[width=\textwidth]{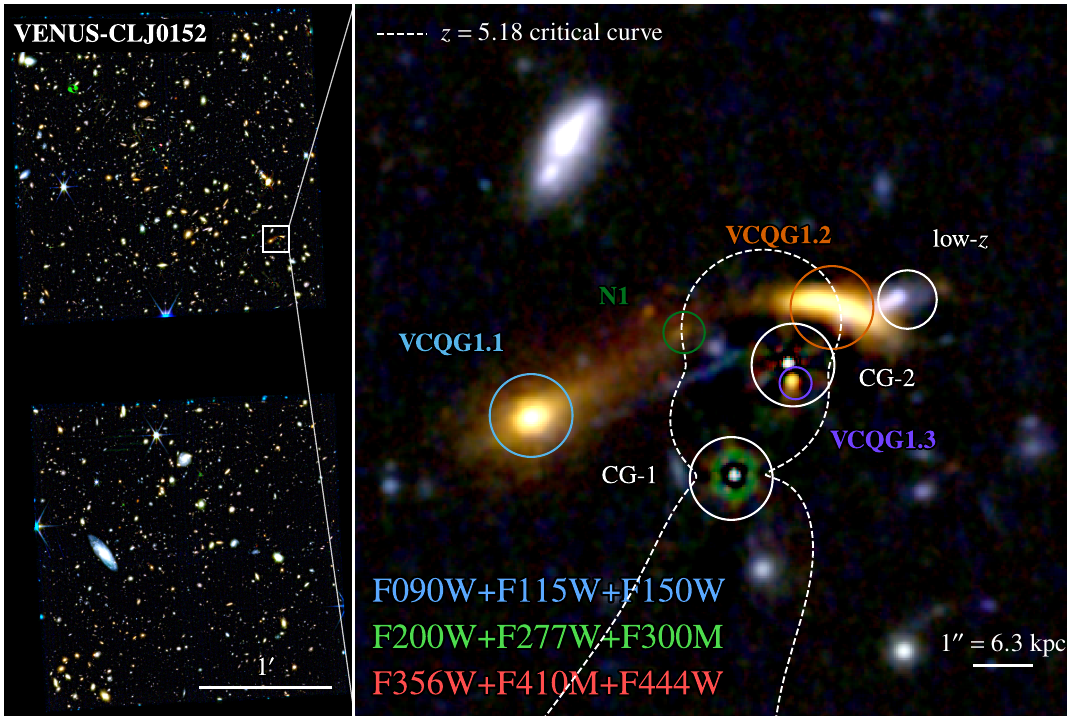}
	\caption{The multiply imaged massive quiescent galaxy VENUS-CLJ0152-QG1 at $z = 5.18$ behind the foreground lensing cluster CL\,J0152.7-1357. Bright cluster galaxy (bCG)-subtracted RGB image of the cluster (left) from \jwst/NIRCam, with F090W+F115W+F150W in blue, F200W+F277W+F300M in green, and F356W+F410M+F444W in red, and a zoom on the system (right). Colored circles mark the images VCQG1.1, VCQG1.2, VCQG1.3 and the neighboring clump N1, white circles mark the cluster members CG-1 and CG-2 and the low-$z$ contaminant, and the dashed line is the $z=5.18$ critical curve. The system comprises three images of VENUS-CLJ0152-QG1 (VCQG1) magnified by the foreground cluster ($z=0.82$). VCQG1.1 is the primary image, initially identified as a Lyman-break candidate in the RELICS survey \citep{coe_relics_2019, salmon_relics_2020}. VCQG1.2 is an elongated secondary image partially blended with a low-$z$ star-forming contaminant, and VCQG1.3 is a demagnified third image blended with the cluster member CG-2. The cluster members CG-1 and CG-2 \citep{Demarco2005} were modeled and subtracted (Section~\ref{subsec:data}). Matching spectral shapes, consistent redshifts, and lensing geometry (Section~\ref{subsec:lens-modelling}) show that VCQG1 is triply imaged.}
	\label{fig:field}
\end{figure*}

\begin{figure*}[t!]
	\centering
	\includegraphics[width=\textwidth]{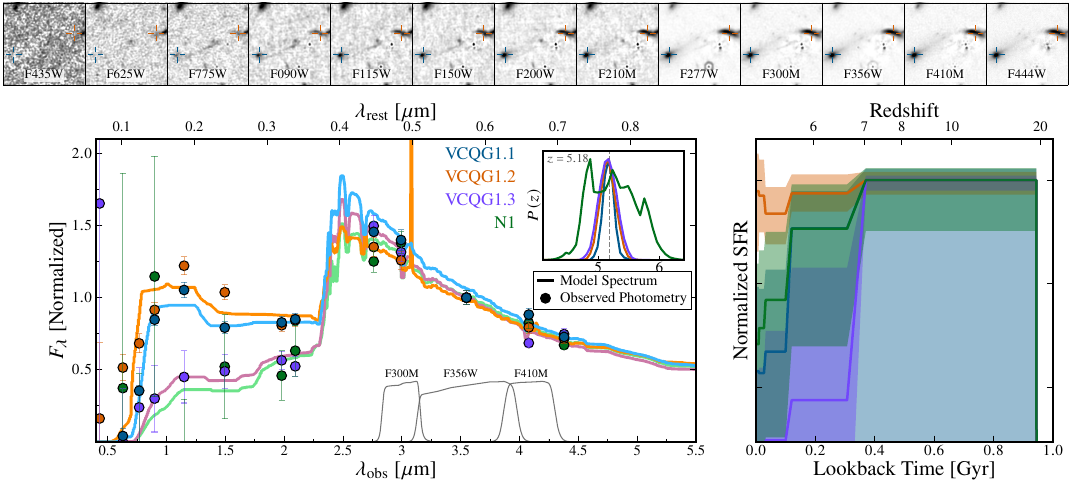}
	\caption{Per-filter stamps, spectral energy distributions and star-formation histories of VCQG1 and N1. The top row shows the 13 single-band stamps of the system, from the \textit{HST}/ACS filters F435W, F625W and F775W to the \jwst/NIRCam filters F090W to F444W, with cross-hairs marking VCQG1.1 and VCQG1.2 in the colors of the lower panels. The bottom left panel shows the SEDs of VCQG1.1, VCQG1.2, VCQG1.3 and N1 normalized at 3.56~$\mu$m (F356W), showing the observed photometry (circles) and the \pros model spectra (lines), together with the transmission curves of the F300M, F356W and F410M filters. The inset shows the redshift probability distribution $P(z)$ from \eazy SED modeling (dashed line, $z=5.18$). The bottom right panel shows the normalized star-formation histories (SFHs), which come from fits to the whole image, not to the central voxels. The neighboring clump N1 has a similar SED and redshift distribution (Section~\ref{subsec:environment}).}
	\label{fig:cutouts}
\end{figure*}

We use the \jwst/NIRCam imaging from the cycle 4 VENUS program (\jwst-GO-6882, PIs: Fujimoto \& Coe, Fujimoto et al. in prep.), which provides 10 photometric bands (F090W, F115W, F150W, F200W, F210M, F277W, F300M, F356W, F410M, and F444W) for 60 lensing clusters including CL\,J0152.7-1357 ($z=0.82$, Figure~\ref{fig:field}). We supplement this with ancillary \textit{HST}/ACS imaging in F435W from the RELICS Survey (GO-14096, \citealt{coe_relics_2019}) and in F625W and F775W from the ACS Guaranteed Time Observations program GTO-9290 (PI Ford), resulting in a 13-band dataset with a 5$\sigma$ point-source depth of $m_{\text{AB}} \approx 29.0$ in F444W. The medium bands F300M and F410M capture the [O~III]+H$\beta$ and H$\alpha$+[N~II] emission-line complexes at $z \approx 5.18$ (Figure~\ref{fig:cutouts}), which constrains the star formation and helps identify emission-line companions in the surrounding field.

All imaging was reduced using the \grizli pipeline \citep{grizli} to a common pixel scale of $0\farcs03$ \citep{weaver_uncover_2024}. To mitigate contamination from bright foreground cluster members, we subtracted the models of neighboring cluster galaxies, including CG-1 and CG-2, and intra-cluster light following \citet{weaver_uncover_2024, atek_jwsts_2025}. We then PSF-matched the mosaics to F444W with empirical PSFs and detected sources on an inverse-variance weighted stack of F277W, F356W, and F444W using \texttt{SourceExtractor} \citep{bertin_sextractor_1996}. The construction and detailing of the photometric catalog will be publicly available in DR1 (Kokorev et al. in prep.). A low-redshift blue interloper in the north-west of VCQG1.2 affects mainly the rest-frame UV continuum (\textit{HST} F435W through \jwst F115W), while the long-wavelength light remains dominated by VCQG1. We mask its pixels after SED modeling.

To preserve spatial information at the $S/N$ needed for SED modeling, we apply Voronoi binning \citep{cappellari_adaptive_2003} to the F444W mosaic within the segmentation map of VCQG1, targeting $S/N \approx 20$ per bin with a minimum of $S/N > 7$. The resulting bins are propagated across all bands to ensure consistent colors and conserved photometry of the de-lensed image. Binning is agnostic to color and could blend similar stellar populations, but the small minimum bin size (4 pixels) limits this effect.

\subsection{SED modeling with EA$z$Y and Prospector}
\label{subsec:sed}

Initial photometric redshifts come from the \textsc{Python} implementation of \eazy \citep{brammer_eazy_2008} with the \textsc{blue\_sfhz\_13} templates, which use redshift-dependent dust attenuation and log-normal SFHs that do not predate the age of the Universe \citep{blanton_k-corrections_2007}. Fitting the PSF-matched photometry in $D=0\farcs{2}$ apertures corrected to total light, we find a maximum \textit{a posteriori} (MAP) redshift of $z=5.18$ for VCQG1.1. We adopt $z=5.18$ as the fixed redshift for all spatially resolved SED modeling, so that the stellar population properties are not degenerate with small redshift fluctuations across the source plane. Fits at $\pm1\sigma$ in redshift give stellar masses and SFRs consistent within $1\sigma$.

We infer the stellar populations with \pros \citep{johnson_stellar_2021}, using a non-parametric continuity SFH prior, \textsc{fsps} \citep{conroy_propagation_2009}, the MILES spectral library \citep{sanchez-blazquez_medium-resolution_2006}, and MIST isochrones \citep{choi_mesa_2016, dotter_mesa_2016}. The SFH has seven bins, with the most recent 100~Myr finely sampled (10, 20, and 70~Myr) to catch rapid quenching and the remaining four bins logarithmically spaced to the Big Bang. We assume the dust attenuation law of \citet{kriek_dust_2013}, with $A_V$ in [0, 2.5], a dust index in [-1, 0.4], and young stars ($t<10^7$~yr) attenuated twice as much as older stars. Gas-phase and stellar metallicities vary over [0.01$Z_{\odot}$, 3$Z_{\odot}$] and [0.1$Z_{\odot}$, 2$Z_{\odot}$], respectively, and $\log{U}$ over [-4, -1]. Posteriors are sampled with \dynesty \citep{speagle_dynesty_2020}, and all properties are corrected for magnification with the lens model (Section~\ref{subsec:lens-modelling}).

Ages and SFHs from photometry alone depend on the priors and on the degeneracies between age, dust, and metallicity \citep{weibel_rubies_2025, de_graaff_efficient_2025}, so we restrict our interpretation to the stellar mass, anchored by the rest-frame optical continuum, and the recent SFR, constrained by the spectral break. Dust geometry or ``outshining'' by recent star formation can still bias both \citep{siegel_uncover_2025, settonUNCOVERNIRSpecPRISM2024}, and the full posteriors are shown in Figure~\ref{fig:corner}.

\subsection{Lens modeling}
\label{subsec:lens-modelling}

We use the updated lens model of CL\,J0152.7-1357 built with the parametric software \texttt{AstroLensPy} \citep{Allingham26a}, a \textsc{Python} implementation of the method of \citet{Zitrin2015} with significant modifications (see also \citealt{pascale22, Furtak2022UNCOVER, Golubchik2025_LRD_VENUS}). The model is constrained by 24 multiply imaged systems, 11 from the \textit{HST} RELICS and \jwst SLICE surveys \citep{Acebron2019, Cerny2026} and 13 newly identified in the 10 NIRCam bands of VENUS, and is described in detail in \citet{allinghamVENUSCensusStrong2026}. The cluster is an ongoing merger along a north-east to south-west axis \citep{Grillo2008, Acebron2019}, and the new multiple images across the field resolve at least three main dark-matter halos along this axis, each modeled as a smooth PIEMD potential \citep{KassiolaKovner1993}, with the cluster members scaled through the Faber-Jackson relation \citep{FaberJackson}. The model reproduces the observed positions of the multiple images with an rms of $0\farcs84$.

VCQG1, discovered by \citet{salmon_relics_2020} and used as a lensed system by \citet{Cerny2026}, lies in the south-western component, while most constraints, including the only spectroscopic redshift, lie in the north-eastern one. We constrain this region with four systems with secure VENUS photometric redshifts and by fixing the relative magnifications of VCQG1.1, VCQG1.2, and VCQG1.3 to their flux ratios, measured within the segmentation map, with the redshift of VCQG1 fixed to $z=5.18$. These constraints largely break the mass-sheet degeneracy, and we impose a 20\% floor on the magnification uncertainty to account for systematics. The model reproduces the three images of VCQG1 (Figure~\ref{fig:lens}), and its magnifications are listed in Table~\ref{tab:detections}.

\subsection{Two-dimensional maps and radial profiles}
\label{subsec:surface density}

Because of the varying number of pixels per Voronoi bin, we normalize the galaxy properties for brightness, stellar mass, and SFR by their physical area ($A$, in kpc$^{2}$). We measure physical areas and distances assuming conservation of surface brightness, $A_{source} \approx A_{image}/\mu$ per pixel and $l_{source}\approx l_{image}/\mu$, in the low-deflection regime.

The light and stellar mass peaks (Figure~\ref{fig:surface_density_profiles}) are measured in the de-lensed plane. Because the strong, asymmetric magnification makes S\'ersic fitting with \galfit or \pysersic \citep{peng_detailed_2002, pasha_pysersic_2023} impractical, we fit radial profiles to the F444W surface brightness and to the resolved stellar mass map, which gives circularized effective radii of the rest-frame optical light and of the stellar mass. Repeating this with the emission-line-free F356W filter gives consistent radii and peaks.

We build color-based emission-line excess maps (Figure~\ref{fig:color_excess}) by subtracting the F356W continuum from F300M and F410M, which contain [O~III]+H$\beta$ and H$\alpha$+[N~II] at $z=5.18$ (Figure~\ref{fig:cutouts}), following \citet{lebowitz_jwst_2025, lorenzMeasuringEmissionLines2025}. These are not equivalent widths, so we interpret only relative brightnesses. An F200W$-$F356W map of the UV and optical continua shows no offset clumps or strong spatial variations, which confirms an unbiased baseline for the emission maps.

The rest-frame UV and optical color maps (Figure~\ref{fig:synth_colors}) come from \eazy with the templates of Section~\ref{subsec:sed} and are compared with an \textsc{FSPS} track for an exponentially declining SFH ($\tau = 200$~Myr) at solar metallicity without dust \citep{fsps2009}. The scatter of VCQG1.2 toward bluer UV colors comes from the low-$z$ contaminant biasing the colors (Figure~\ref{fig:field}). Constraints on the rest-frame $J$ and synthetic $i$ bands are limited by dust uncertainties beyond the NIRCam wavelength coverage, and strong dust gradients in the cores could be present \citep{siegel_uncover_2025, settonUNCOVERNIRSpecPRISM2024}.

\section{Analysis} \label{sec:results}

\subsection{A multiply imaged massive quenched galaxy}
VCQG1 was originally identified as a Lyman-break candidate at $z \sim 5$ in the \textit{Hubble Space Telescope} (\textit{HST}) Reionization Lensing Cluster Survey (RELICS) \citep{salmon_relics_2020}. Our \eazy fits of the three images give consistent MAP redshifts of $z=5.18$, $5.16$, and $5.15$ for VCQG1.1, 1.2, and 1.3, and their similar breaks at $\lambda_{\text{rest}}\approx 4000$\,\AA\ (Figure~\ref{fig:cutouts}) suggest that they are multiple images of one galaxy, a likely post-starburst system dominated by an evolved population of late-B and A-type stars. The \pros SED models agree across the images, except for the low UV continuum of VCQG1.3, which is significantly blended with the subtracted cluster galaxy CG-2. The lens model independently supports the triply imaged solution (Figure~\ref{fig:lens}), with area-averaged magnifications of $\mu \approx 6.7$, $10.7$, and $0.74$ (Table~\ref{tab:detections}).

For VCQG1.1 and VCQG1.2, we infer stellar masses of $\text{M}_{*} = 3.36^{+0.64}_{-0.47} \times 10^{10}$ and $2.77^{+0.59}_{-0.42}\times 10^{10}\,\text{M}_{\odot}$, respectively. The low star-formation rates (SFR) averaged over the last 30~Myr are $\text{SFR}_{30} = 4.07^{+1.65}_{-0.61}$ and $4.63^{+3.70}_{-1.75}\, \text{M}_{\odot}\,\text{yr}^{-1}$. The resulting specific star-formation rates of $\text{sSFR}_{30} = 1.21^{+0.22}_{-0.01}\times 10^{-10}$ and $1.67^{+0.81}_{-0.45}\times 10^{-10}\,\text{yr}^{-1}$ correspond to mass-doubling times of $8.24^{+0.10}_{-1.25}$ and $5.99^{+2.19}_{-1.95}$~Gyr at constant SFR, well above both the age of the Universe at $z=5.18$ ($\sim1.12$\,Gyr) and the $5\,t_{H} = 5.6$\,Gyr threshold that defines quiescence \citep{carnall_massive_2023}, so VCQG1 is recently quenched.

\begin{figure*}[t!]
	\centering
	\includegraphics[width=\textwidth]{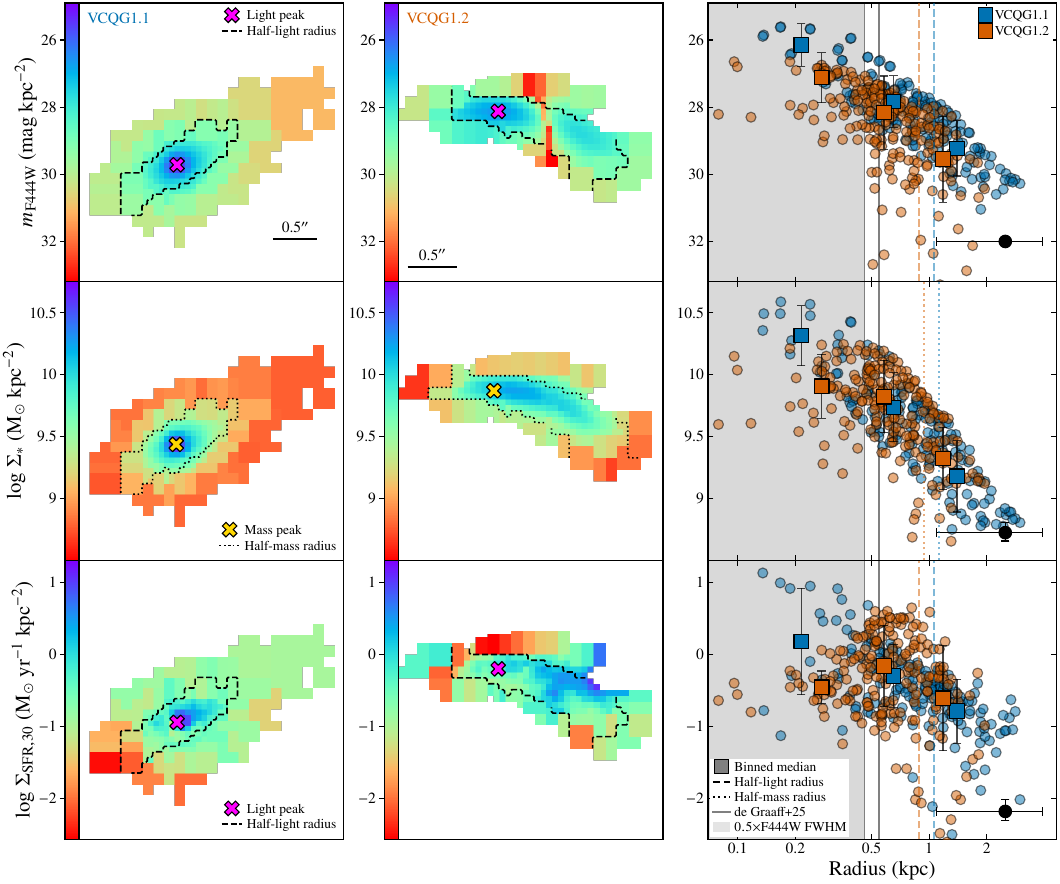}
	\caption{Resolved structural profiles and parameter maps of VENUS-CLJ0152-QG1.
		Maps of the surface brightness in F444W (top), stellar mass surface density ($\Sigma_*$, middle), and
		star-formation rate surface density ($\Sigma_{\text{SFR}}$, bottom) for images VCQG1.1 (left column) and VCQG1.2 (middle column).
		The right column shows radial profiles of the same quantities for both images. Squares are binned medians, and black circles
		show typical uncertainties. The grey shaded region marks 0.5 times the full width at half maximum (FWHM) of the NIRCam
		F444W point spread function ($\sim0.46$~kpc), below which structures are unresolved without lensing. Dashed and dotted lines mark the half-light
		and half-mass radii, R$_{\text{eff, F444W}} = 1.06 \pm 0.01$ and $0.88 \pm 0.01$~kpc, and
		R$_{\text{eff, *}} = 1.13 \pm 0.02$ and $0.94 \pm 0.02$~kpc for VCQG1.1 and VCQG1.2, respectively. The solid grey
		line marks the effective radius of REQG1, a $z \sim 5$ quiescent galaxy in an unlensed field
		\citep{de_graaff_efficient_2025}. The $\Sigma_{\text{SFR}}$ maps show a localized enhancement (blue pixels) offset
		from the light peak (magenta cross). The two profiles agree within $1\sigma$ given lens-model and photometric
		uncertainties.}
	\label{fig:surface_density_profiles}
\end{figure*}

\subsection{An extended, low-density stellar structure}
From the two-dimensional surface density maps, we measure half-light and intrinsic half-mass radii of R$_{\text{eff, F444W}} = 1.06 \pm 0.01$ and $0.88 \pm 0.01$~kpc, and R$_{\text{eff, *}} = 1.13 \pm 0.02$ and $0.94 \pm 0.02$~kpc for VCQG1.1 and VCQG1.2, respectively (Figure~\ref{fig:surface_density_profiles}). The central stellar mass surface densities within 1~kpc are $\log(\Sigma_{*, \text{1kpc}}/M_\odot\,\text{kpc}^{-2}) = 9.86 \pm 0.01$ and $9.83 \pm 0.01$. Within the effective radius, $\Sigma_{e}=(M_*/2)/(\pi R_{e}^2)$ with $R_e$ the circularized half-mass radius \citep{barro_structural_2017}, we find $\log\Sigma_e=9.62^{+0.08}_{-0.07}$ and $9.70^{+0.08}_{-0.07}$ (Figure~\ref{fig:sigma1}).

\begin{figure*}[t!]
	\centering
	\includegraphics[width=\textwidth]{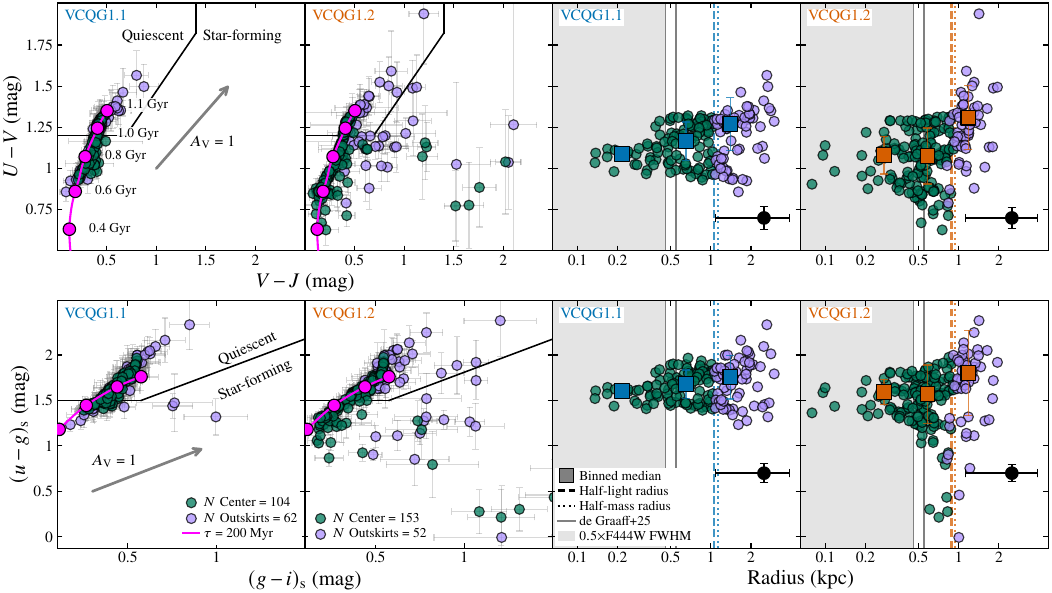}
	\caption{Spatially resolved rest-frame UV versus optical colors and the UV radial gradients of
		VENUS-CLJ0152-QG1.
		The left two columns show \eazy modeled rest-frame $UVJ$ (top) and synthetic $ugi$ (bottom) color--color
		distributions for VCQG1.1 (left) and VCQG1.2 (second column).
		The right two columns show radial rest-frame UV profiles, $U-V$ (top) and $(u-g)_s$ (bottom), for VCQG1.1 (third column) and
		VCQG1.2 (right column). Panels in the same row share the same y axis.
		Black circles mark inner core spaxels ($r < R_{\text{eff, F444W}}$), yellow circles denote outer spaxels
		($r \geq R_{\text{eff, F444W}}$), and colored squares show binned radial medians. The grey shaded region in the right two columns marks 0.5 times the F444W FWHM
		(Figure~\ref{fig:surface_density_profiles}). The solid wedge delineates
		quiescent from star-forming galaxies \citep{antwi-danso_beyond_2023}, and the black arrow in the top left panel indicates an extinction
		vector of $A_V = 1$.
		For VCQG1.2, there is significant scatter toward bluer UV and redder optical colors, largely due to the blended
		low-$z$ star-forming contaminant. A magenta track illustrates the color evolution for a simple stellar population
		model with an exponentially declining star-formation history [$\tau = 200$~Myr, \citealp{pythonfsps2014}]. This is
		consistent with the rapid formation and abrupt quenching found in $z>4$ analogs
		\citep{carnall_massive_2023, weibel_rubies_2025, de_graaff_efficient_2025}. The resolved UV colors exhibit positive
		gradients and prohibit potential uniform or inside-out quenching pathways which produce flat or negative color
		gradients, respectively \citep{edward_magaz3ne_2026}.}
	\label{fig:synth_colors}
\end{figure*}

\subsection{A younger core revealed by resolved colors}
Because color gradients trace the evolutionary pathway \citep{edward_magaz3ne_2026}, we measure rest-frame UV ($U-V$ and $(u-g)_s$) and optical ($V-J$ and $(g-i)_s$) colors with \eazy, whose linear template combinations match the observed photometry more closely than the physically motivated \pros models. Both inside ($r < \text{R}_{\text{eff}}$) and outside the effective radius, the colors follow the track of the $\tau = 200$~Myr model \citep{pythonfsps2014} (Figure~\ref{fig:synth_colors}). Despite the degeneracies between age, dust, and metallicity, the radial profiles show clear positive color gradients, indicative of a younger, bluer core and redder outskirts. This independently supports a recently quenched, post-starburst galaxy in which the core quenched only recently.

\begin{figure*}[t!]
	\centering
	\includegraphics[width=\textwidth]{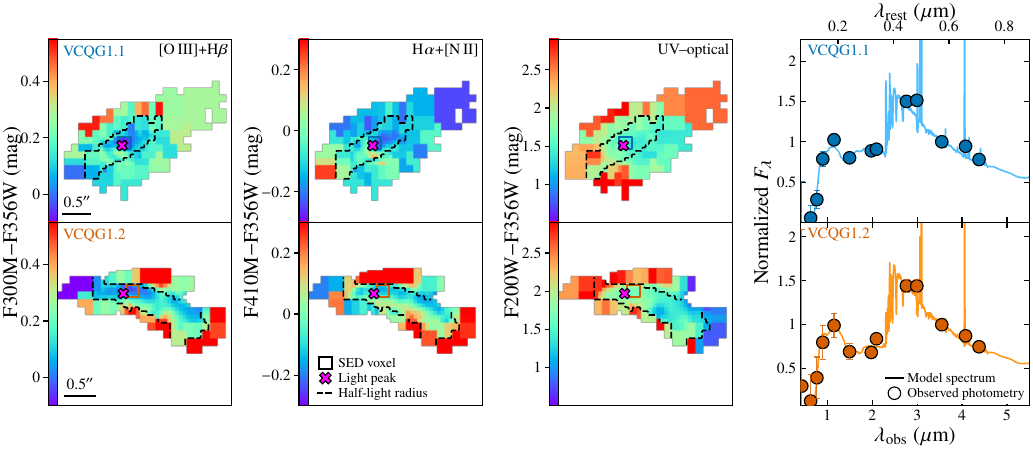}
	\caption{Spatially resolved color excess and emission-line maps of VENUS-CLJ0152-QG1.
		VCQG1.1 (top) and VCQG1.2 (bottom). At $z=5.18$, medium bands F300M (first column) and F410M (second column) isolate
		rest-frame $[\text{O}\,\text{III}]\lambda\lambda4959,5007+\text{H}\beta$ and
		$\text{H}\alpha+[\text{N}\,\text{II}]\lambda\lambda6548,6583$, respectively, relative to the F356W optical continuum.
		The F200W (third column) isolates the UV continuum. The right column shows \pros modeled SEDs from selected voxels
		with brighter line emissions (open squares in the first three columns). Bluer tones in the color maps indicate brighter line
		excess. Magenta crosses mark the light centroid, and dashed contours trace the half-light radius. Both
		emission-line excesses coincide with the off-center clump located $\sim$0.7~kpc from the luminous core, mirroring the
		spatial offset mapped in Figure~\ref{fig:surface_density_profiles}. In unlensed observations, this feature would blend
		into the stellar core and mimic a central starburst or central active galactic nucleus. Instead, resolving this
		spatial separation points to alternative evolutionary pathways only viewable with lensing boosts.}
	\label{fig:color_excess}
\end{figure*}

\subsection{Off-center ionized gas emission}
The central region appears to retain residual star formation with a SFR surface density of $\Sigma_{\text{SFR}_{30, 1\,\text{kpc}}}=1.08^{+0.05}_{-0.04}\,M_\odot\,\text{yr}^{-1}\text{kpc}^{-2}$ (Figure~\ref{fig:surface_density_profiles}, bottom row). This translates to a low sSFR of $\text{sSFR}_{30, 1\,\text{kpc}}= 1.48^{+0.71}_{-0.53}\times 10^{-10}\,\text{yr}^{-1}$, which is well below the quenching threshold. Strikingly, the two-dimensional SFR surface density map reveals that the star-formation is concentrated in a clump physically offset from the central peak. The emission-line maps (Section~\ref{subsec:surface density}) show strong F300M$-$F356W and F410M$-$F356W color excesses, which trace [O~III]+H$\beta$ and H$\alpha$+[N~II] emission at the same position, about 0.7~kpc from the stellar core (Figure~\ref{fig:color_excess}). This suggests that the inferred star-formation enhancement is driven almost entirely by these color excesses, and therefore by [O~III]+H$\beta$ and H$\alpha$+[N~II] emission. Because the F200W$-$F356W continuum map shows no such offset, we conclude that the enhancement is unique to F300M and F410M (Figure~\ref{fig:color_excess}, third column).

\subsection{Environment and neighboring populations}
\label{subsec:environment}

Based on \eazy photometric redshifts and lens modeling, we identify one potential neighboring galaxy (N1 in Figure~\ref{fig:field}) at $z=5.4\pm{0.6}$ within a projected source-plane distance of $\sim 10$~kpc if residing at the same redshift. Its SED shows a similar break between F200W and F277W, and its $p(z)$ overlaps strongly with those of the three images of VCQG1, which are centered at $z=5.18$ (Figure~\ref{fig:cutouts}).

\begin{figure*}[t!]
	\centering
	\includegraphics[width=\textwidth]{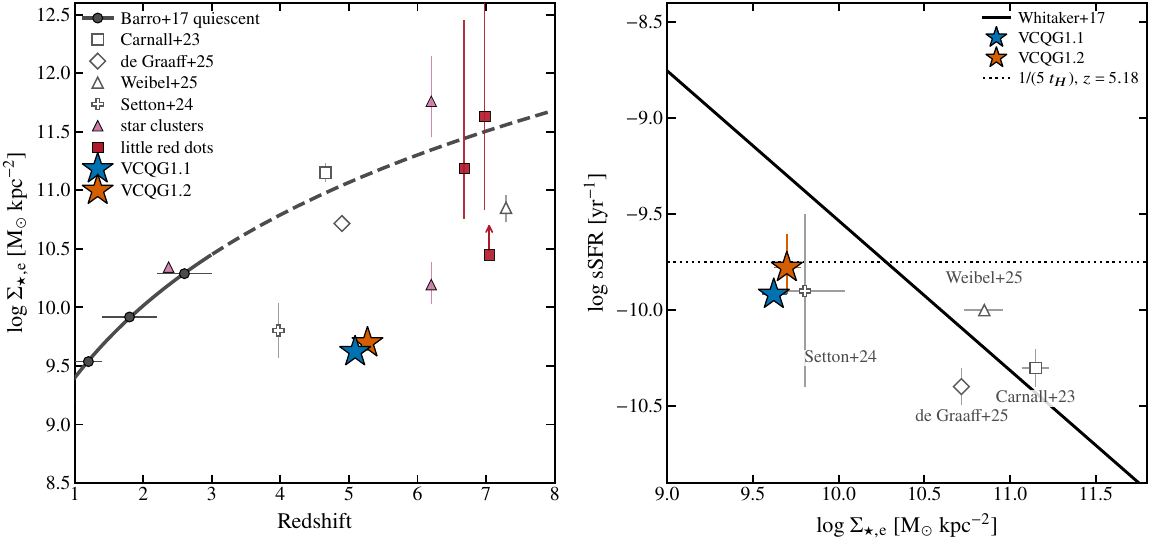}
	\caption{Stellar mass surface density within the effective radius of VCQG1 compared with quiescent galaxies. The left panel shows $\log\Sigma_e$ as a function of redshift for quiescent galaxies at $z>3$ \citep{carnall_massive_2023, de_graaff_efficient_2025, weibel_rubies_2025, settonUNCOVERNIRSpecPRISM2024} and for quiescent galaxies with the stellar mass of VCQG1 at $0.5<z<3$ from \citet{barro_structural_2017} (dark grey points at the bin centers, with horizontal bars marking the bin widths). The dark grey curve is a fit of $\log\Sigma_e=8.34+3.50\log(1+z)$ to these points, solid over the range of the data and dashed where extrapolated beyond $z=3$. For comparison, colored symbols show non-quiescent compact systems at high redshift: young massive star clusters (pink triangles) in the Sunburst arc \citep{vanzella_high_2022}, the Sunrise arc \citep{vanzella_jwstnircam_2023} and the Firefly Sparkle \citep{mowla_formation_2024}, and little red dots (red squares) from \citet{furtak_high_2024} and \citet{baggen_small_2024}. For these, $\Sigma_e$ is computed with the same definition from the published stellar masses and half-light radii. Arrows mark lower limits, and the error bars of the \citet{baggen_small_2024} little red dots span their range of assumed AGN contributions. The right panel shows $\log\text{sSFR}$ as a function of $\log\Sigma_e$, with the relation of galaxies at $1.5<z<2.5$ from \citet{whitaker_predicting_2017} and the quiescence threshold $\text{sSFR}=1/(5\,t_H)$ at $z=5.18$ ($\log\text{sSFR}=-9.75$). VCQG1.1 and VCQG1.2 are shown as blue and orange stars, with $\log\Sigma_e=9.62^{+0.08}_{-0.07}$ and $9.70^{+0.08}_{-0.07}$, respectively. $\Sigma_e=(M_*/2)/(\pi R_e^2)$ with $R_e$ the circularized half-mass radius.}
	\label{fig:sigma1}
\end{figure*}

\section{Discussion} \label{sec:discussion}

\subsection{Central density and quiescence}
\label{subsec:disc_density}
VCQG1 is among the most massive galaxies known at $z>4$ \citep{carnall_massive_2023, antwi-danso_feniks_2025,de_graaff_efficient_2025, weibel_rubies_2025} and lies above the knee of the stellar mass function at $z\sim5$ [$\sim2.0 \times 10^{10}M_{\odot}$, \citealp{weaverCOSMOS2020GalaxyStellar2023}]. Yet it is $\sim 2$ to $4$ times more extended than similar quiescent galaxies at $z>4$ \citep{de_graaff_efficient_2025, pascalauWhenRelicsWere2026, weibel_rubies_2025}, its $\Sigma_{*,\text{1kpc}}$ is about 1 dex lower than in field-selected analogs \citep{ji_reconstructing_2023, de_graaff_efficient_2025, weibel_rubies_2025}, and is comparable only to that of \citet{settonUNCOVERNIRSpecPRISM2024} at $z=3.97$ (Figure~\ref{fig:sigma1}, left). Part of this gap may come from comparing a resolved galaxy with unresolved ones. Many field systems have sizes of $\sim 0.2$ to $0.6$ kpc, comparable to the NIRCam F444W PSF limit ($\approx 0.46$ kpc at these redshifts), and unresolved profiles risk extrapolating high core densities across the galaxy, inflating the inferred stellar mass and age \citep{cutler_unbreaking_2026}. Lensing avoids this blurring bias, and resolved measurements of more quiescent galaxies at $z>4$ will show how much of this gap is intrinsic.

Figure~\ref{fig:sigma1} (right) places VCQG1 in the $\Sigma_{e}$--sSFR plane, with the relation of galaxies at $1.5<z<2.5$ from \citet{whitaker_predicting_2017} and the quiescence threshold $\text{sSFR}=1/(5\,t_H)\approx1.8\times10^{-10}\,\text{yr}^{-1}$ at $z=5.18$. VCQG1 lies 0.5 to 0.7 dex below the relation, so it is more quenched than expected for its central density, with VCQG1.1 and VCQG1.2 below the threshold within their uncertainties. Its $\Sigma_e$ matches quiescent galaxies of the same stellar mass at $1.0<z<1.4$ from \citet{barro_structural_2017} and is 0.2 to 0.7 dex lower than at $1.4<z<3.0$, whereas extrapolating their evolution to $z=5.18$ predicts $\log\Sigma_e\approx11.1$, close to the field quiescent galaxies at $z>4$ and 1.4 to 1.5 dex above VCQG1. The extended profile and moderate central density of VCQG1 suggest that it quenched without compaction, which challenges the compaction-driven quenching sequences of recent theoretical models \citep{saputraharyanaInterplayCompactionQuenching2026, kimmig_blowing_2025, ni_astrid_2025}. Its $\Sigma_1$ matches that of quiescent galaxies at Cosmic Noon \citep{barro_structural_2017, whitaker_predicting_2017, moslehConnectionStellarMass2017}. Because it is already quiescent, it is unlikely to grow much further in stellar mass, which makes systems like VCQG1 plausible direct progenitors of the Cosmic Noon population.

\subsection{Colors and radial gradients}
\label{subsec:disc_colors}
The positive color gradients (Figure~\ref{fig:synth_colors}) rule out uniform or inside-out quenching and ex-situ minor mergers, which produce flat or negative gradients \citep{edward_magaz3ne_2026}, and show a galaxy caught shortly after quenching, before it becomes indistinguishable from older relaxed systems \citep{whitaker_quenching_2026}. Quiescent galaxies at Cosmic Noon instead have older, metal-rich cores and bluer outskirts [e.g., \citealp{suess_color_2020, cheng_age_2024}], so VCQG1 must reverse its gradient to become one of them. Continued accretion of younger, lower-mass, metal-poor companions such as N1 would deposit fresh stars in its halo \citep{naab_minor_2009, vandokkumFormingCompactMassive2015} and could turn the positive gradient negative, linking early quenched galaxies to their Cosmic Noon descendants.

\subsection{The offset emission clump}
\label{subsec:disc_clump}
The ionized gas emission sits $\sim0.7$ kpc from the stellar core (Figure~\ref{fig:color_excess}), which makes its origin a direct test of central AGN quenching at these redshifts. In an unlensed field, blurring would merge it with the core and mimic the central starbursts or AGN seen in high-redshift analogs [e.g., \citealp{carnall_massive_2023,de_graaff_efficient_2025, weibel_rubies_2025}]. Line ratios from spectroscopy are needed to separate AGN and stellar photoionization \citep{kewley_theoretical_2013}, but the maps and SED show no clear central unobscured AGN. We consider two possible physical origins for the offset emission.

\textit{1) AGN outflow or jet shock-ionization.} Simulations and observed analogs require a supermassive black hole to heat or eject gas to quench star formation in the early Universe \citep{kimmig_blowing_2025, ni_astrid_2025, alexander_what_2025, de_graaff_efficient_2025, weibel_rubies_2025}. In VCQG1, the lack of a central AGN signature in the emission lines or a rising optical continuum would then imply a recently switched-off or heavily obscured AGN \citep{hickox_obscured_2018, alexander_what_2025}, with the offset clump tracing an outflow or jet-driven shocks \citep{revalski_quantifying_2025}. Obscured AGN with ionized outflows in evolved hosts at $z>5$ point to radiative feedback \citep{deugenioJADESSAPPHIRESGalaxy2025, deugenioDustyQuenchingcandidateAGN2026}, while radio jets shock-ionizing dusty gas point to kinetic feedback \citep{duncan_jwsts_2023, roy_jwst_2024, saxena_widespread_2024}. Because lensing resolves the emission $\sim0.7$\,kpc from the stellar core, future potential resolved spectroscopy of VCQG1 could measure the kinematics and line ratios of the clump apart from the core and tell these two modes apart.

\textit{2) A merger with N1.} A merging companion could infuse the system with cold gas and trigger a localized burst of star formation. N1, a possibly stripped clump between VCQG1.1 and VCQG1.2 at $z = 5.4 \pm 0.6$ (Section~\ref{subsec:environment}), is a potential candidate. This late rejuvenation fits the bursty star formation of early galaxies \citep{faisstDeadAliveHow2024,remusRelightCandleWhat2025} and would explain both the positive color gradient, with its younger, bluer core, and the offset clump with its ionized gas emission. Its low SFR adds little stellar mass, consistent with little later mass growth and with VCQG1 being a progenitor of Cosmic Noon MQGs. Models predict that mergers trigger compaction and a central starburst, yet VCQG1 is extended, moderately dense, and passive, so if indeed confirmed, this post-merger scenario would show that merger-driven quenching can proceed without compaction, a pathway not yet expected at such high redshifts.

\subsection{Caveats and future work}
\label{subsec:caveats}
Our analysis relies on the photometric redshift of VCQG1. Its three images give consistent \eazy redshifts ($z=5.15-5.18$) that agree with the lensing geometry, and we fix $z=5.18$ in both the lens model and the SED fits. Repeating the fits at $\pm1\sigma$ in redshift recovers the stellar masses and SFRs within $1\sigma$ (Appendix~\ref{app:properties}), so our measurements are robust to this choice. \jwst/NIRSpec spectroscopy, ideally with its integral field unit, would remove this assumption and add information that photometry cannot provide.

Such spectra would confirm the redshift and quiescence through the Balmer break and stellar absorption lines, and measure absorption-line ages and metallicities that break the degeneracy between age, dust, and metallicity in the SED fits. Resolved spectroscopy, aided by the magnification, would measure the stellar and gas kinematics to test whether VCQG1 is rotationally supported \citep{newman_discovery_2015, forrestMassiveEvolvedSlowrotating2026}, separate [O~III] from H$\beta$ and H$\alpha$ from [N~II] to measure the line ratios of the offset clump and distinguish star formation, AGN, and shocks, and reveal broad or high-ionization lines from current or obscured AGN activity. A spectroscopic redshift of N1 would test whether it is merging with VCQG1.

VENUS-CLJ0152-QG1 represents a rapidly formed, abruptly quenched early system. It is not the compact, centrally-dense, axi-symmetric galaxy typical of recent observations \citep{antwi-danso_feniks_2025, weibel_rubies_2025, de_graaff_efficient_2025, carnall_massive_2023} or cosmological simulations \citep{chandro-gomez_unveiling_2026, kimmig_blowing_2025}. Our findings show that there may be a diversity in evolutionary pathways that unlensed fields would be incapable of identifying. Resolving spatial structures in these rare lensed systems could alleviate the disparity between theoretical feedback models and observations of massive early galaxies.

\begin{acknowledgments}
The authors acknowledge the use of the Canadian Advanced Network for Astronomy Research (CANFAR) Science Platform
operated by the Canadian Astronomy Data Center (CADC) and the Digital Research Alliance of Canada (DRAC), with support
from the National Research Council of Canada (NRC), the Canadian Space Agency (CSA), CANARIE, and the Canadian
Foundation for Innovation (CFI). This work is based on observations made with the NASA/ESA/CSA \textit{James Webb Space
Telescope} obtained at the Space Telescope Science Institute, which is operated by the Association of Universities for
Research in Astronomy, Incorporated, under NASA contract NAS5-03127. The \jwst data presented in this article
were obtained from the Mikulski Archive for Space Telescopes (MAST) at the Space Telescope Science Institute. These
observations are associated with program \jwst GO-6882 (VENUS). This work also uses \textit{HST}/ACS observations from programs GO-14096 (RELICS) and GTO-9290. The \jwst and \textit{HST} data used in this paper can be found in the Mikulski Archive for Space Telescopes (MAST) at \dataset[doi:10.17909/pxsv-0758]{https://doi.org/10.17909/pxsv-0758} (VENUS), \dataset[doi:10.17909/T9SP45]{https://doi.org/10.17909/T9SP45} (RELICS), and \dataset[doi:10.17909/0v5f-ps18]{https://doi.org/10.17909/0v5f-ps18} (GTO-9290). We acknowledge the use of an AI-assisted tool
(Claude Opus 5.5, Anthropic) to format the visualizations of the results, as well as to refine the language to improve clarity.

S.F. acknowledges support from the Dunlap Institute, funded through an endowment established by the David Dunlap
family and the University of Toronto. We acknowledge the support of the Canadian Space Agency (CSA) [25JWGO4A18].
A.Zi. acknowledges support by the Israel Science Foundation Grant No. 864/23. P.D. warmly acknowledges support from an
NSERC discovery grant (RGPIN-2025-06182). F.E.B. acknowledges support from ANID-Chile BASAL CATA FB210003 and FONDECYT
Regular 1241005. E.V. acknowledges financial support through grants INAF GO 2024 and by the European Union --
NextGenerationEU within PRIN 2022 project n.20229YBSAN. H.Y. and M.N. acknowledge support by KAKENHI (25KJ0832) through
Japan Society for the Promotion of Science (JSPS). V.M. and M.B. acknowledge support from the ERC Grant FIRSTLIGHT
\#~101053208, Slovenian national research agency ARIS through grants N1-0238 and P1-0188. M.B. acknowledges support ESA
PRODEX Experiment Arrangements No. 4000146646 and 4000149972. R.A.W. acknowledges support from NASA JWST
Interdisciplinary Scientist grants NAG5-12460, NNX14AN10G and 80NSSC18K0200 from GSFC.
\end{acknowledgments}

\facilities{JWST(NIRCam), HST(ACS, WFC3)}

\software{Gri$z$li \citep{grizli}, Source Extractor \citep{bertin_sextractor_1996}, VorBin \citep{cappellari_adaptive_2003}, EA$z$Y \citep{brammer_eazy_2008}, Prospector \citep{johnson_stellar_2021}, dynesty \citep{speagle_dynesty_2020}, Astropy \citep{astropy_collaboration_astropy_2022}}

\clearpage
\appendix

\section{Lens model reproduction of VCQG1} \label{app:lens}
\setcounter{figure}{0}
\renewcommand{\thefigure}{A\arabic{figure}}

\begin{figure*}
	\centering
	\includegraphics[width=0.8\textwidth]{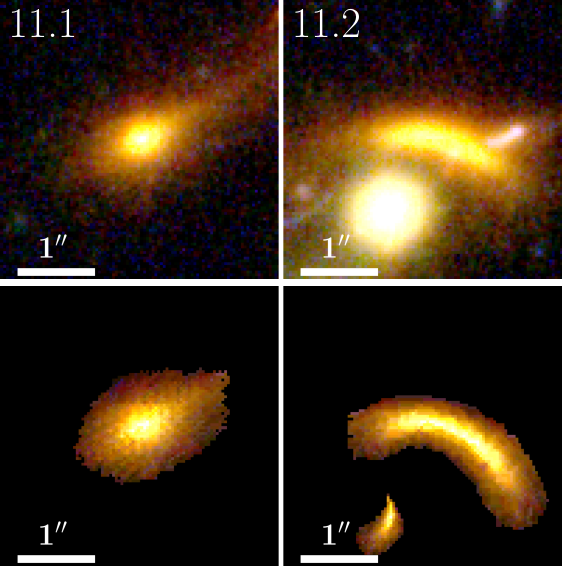}
	\caption{VCQG1 main images, denoted as system 11 in our lens model.
		The top row shows the observed lensed images and the bottom row the reproduced model images. In each case, we de-project image 11.1
		(VCQG1.1) into the source plane, and project it back into the lens plane. The similar position, geometry and fluxes
		demonstrate the quality of the lens model.}
	\label{fig:lens}
\end{figure*}

\section{Derived properties} \label{app:properties}
\setcounter{figure}{0}
\renewcommand{\thefigure}{B\arabic{figure}}
\setcounter{table}{0}
\renewcommand{\thetable}{B\arabic{table}}

\begin{figure*}
	\centering
	\includegraphics[width=\textwidth]{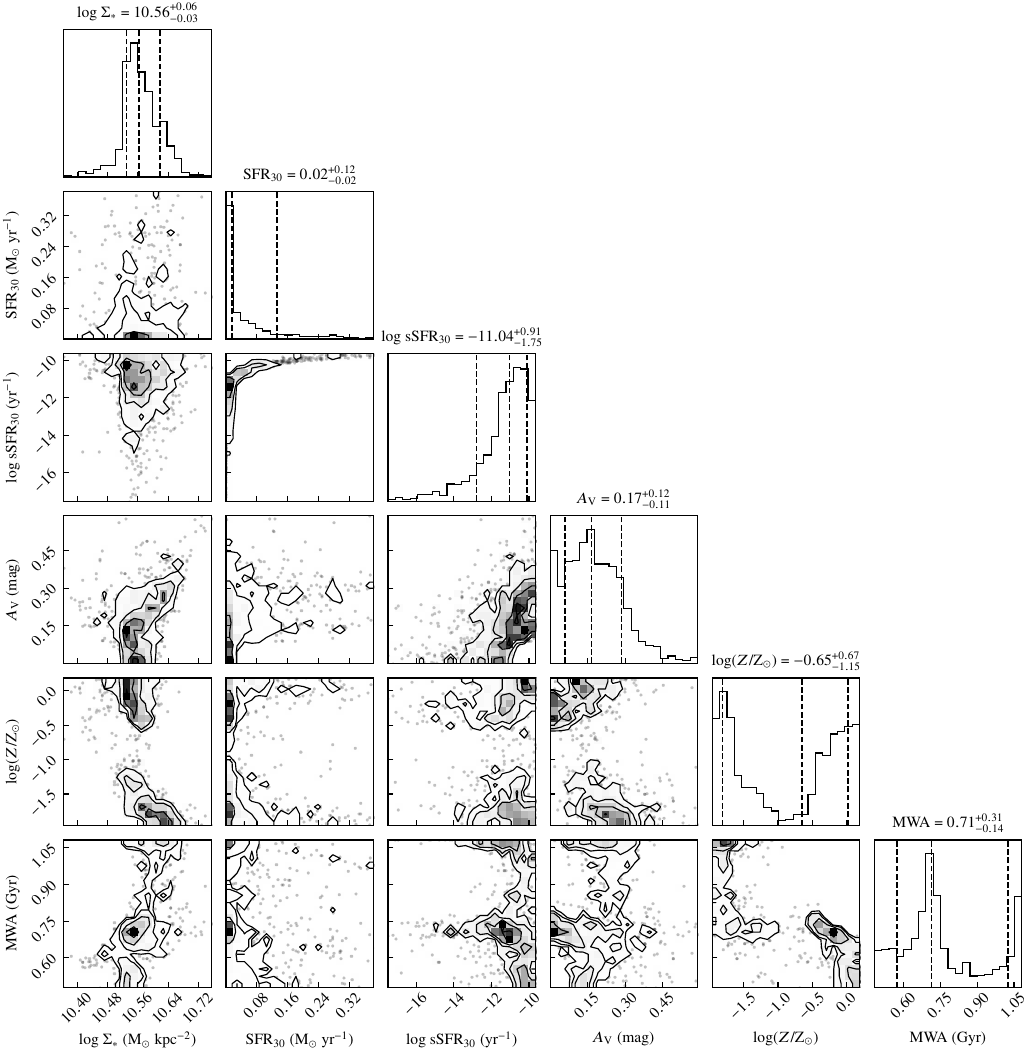}
	\caption{Posterior distributions of the fiducial \pros model.
		Posterior probability distributions for stellar mass surface density, star-formation rate and specific
		star-formation rate averaged over the last 30~Myr, dust extinction, stellar metallicity, and mass-weighted age
		derived from our fiducial \pros model. Off-diagonal panels show two-dimensional joint posteriors, with
		contours enclosing 12\%, 39\%, 68\%, and 86\% of the probability mass. Diagonal panels show
		the one-dimensional marginalized posteriors, where vertical dashed lines indicate the 16th, 50th, and 84th
		percentiles. The
		redshift posterior is not shown because we fixed the redshift to the MAP redshift.}
	\label{fig:corner}
\end{figure*}

Table~\ref{tab:detections} lists the positions, F444W magnitudes, magnifications, redshifts, and magnification-corrected stellar masses and SFRs of the three images of VCQG1, and Figure~\ref{fig:corner} shows the posterior distributions of the fiducial \pros model. Photometry alone cannot fully resolve the uncertainties of the lens model or of the SED model. We include the lens-model uncertainty through a 20\% floor on the magnification, and because the degeneracies between age, dust, and metallicity remain (Figure~\ref{fig:corner}), we interpret only the stellar mass and the recent SFR (Section~\ref{subsec:sed}). To test the fixed redshift, we repeated the fits at $z\pm1\sigma$ and recover the stellar masses and SFRs within $1\sigma$.

\begin{table*}
	\centering
	\caption{Detected images and derived physical properties of VENUS-CLJ0152-QG1. For the highly magnified VCQG1.1 and VCQG1.2 images, we report \pros inferred physical properties derived using spatially resolved analysis. For the demagnified VCQG1.3 image, only aperture photometry is reported. All physical measurements are corrected for local magnification. Columns are (i) Object ID and corresponding methodology; (ii) \& (iii) R.A. and Dec. in degrees; (iv) Observed magnitude in the \jwst/NIRCam F444W filter; (v) Lensing magnification $\mu$; (vi) Redshift inferred using \eazy; (vii) \& (viii) The magnification-corrected stellar mass and star formation rate averaged over the most recent 30~Myr, inferred from \pros SED modeling.
    \newline $^{*}$ To account for systematic effects during the fitting analysis, we impose an error floor of 5\% on the observed photometric uncertainties.
    \newline $^{\dagger}$ Reported values represent the median and $1\sigma$ uncertainties derived from the magnification distributions in the spatially resolved lens modeling.
    \newline $^{\ddagger}$ For the spatially resolved \pros modeling, we fix the redshift to the maximum a posteriori (MAP) value of $z=5.18$ inferred from \eazy SED modeling.}
	\label{tab:detections}
	\resizebox{\textwidth}{!}{%
    \begin{tabular}{l c c c c c c c}
    \hline
    \hline
    Image (Method) & $\alpha$ & $\delta$ & F444W Mag$^{*}$ & $\mu^{\dagger}$ & Redshift$^{\ddagger}$ & $\text{M}_*$ & $\text{SFR}_{30}$ \\
    & (deg, J2000) & (deg, J2000) & (AB mag) & & & ($10^{10}\,\text{M}_\odot$) & ($\text{M}_\odot\,\text{yr}^{-1}$) \\
    \hline
    VCQG1.1 (Resolved) & 28.1656670 & -13.9687161 & 21.635 $\pm$ 0.003 & $6.77^{+0.59}_{-0.41}$ & $5.17^{+0.09}_{-0.09}$ & $3.36^{+0.64}_{-0.47}$ & $4.07^{+1.65}_{-0.61}$ \\
    VCQG1.2 (Resolved) & 28.1642006 & -13.9682098 & 21.516 $\pm$ 0.002  & $10.71^{+20.57}_{-4.97}$ & $5.16^{+0.13}_{-0.12}$ & $2.77^{+0.59}_{-0.42}$ & $4.63^{+3.70}_{-1.75}$ \\
    VCQG1.3 (Aperture) & 28.1644409 & -13.9685689 & 24.865 $\pm$ 0.018 & $0.74 \pm 0.10$ & $5.16^{+0.15}_{-0.15}$ & $1.62^{+0.18}_{-0.16}$ & $0.08^{+0.46}_{-0.08}$ \\
    \hline
    \hline
    \end{tabular}%
    }
\end{table*}

\bibliographystyle{aasjournalv7}
\bibliography{vcqg1_refs}

\begin{thebibliography}{}
\expandafter\ifx\csname natexlab\endcsname\relax\def\natexlab#1{#1}\fi
\providecommand{\url}[1]{\href{#1}{#1}}
\providecommand{\dodoi}[1]{doi:~\href{http://doi.org/#1}{\nolinkurl{#1}}}
\providecommand{\doeprint}[1]{\href{http://ascl.net/#1}{\nolinkurl{http://ascl.net/#1}}}
\providecommand{\doarXiv}[1]{\href{https://arxiv.org/abs/#1}{\nolinkurl{https://arxiv.org/abs/#1}}}

\bibitem[{A. {Acebron} {et~al.}(2019){Acebron}, {Alon}, {Zitrin}, {Mahler},
  {Coe}, {Sharon}, {Cibirka}, {Brada{\v{c}}}, {Trenti}, {Umetsu},
  {Andrade-Santos}, {Avila}, {Bradley}, {Carrasco}, {Cerny}, {Czakon},
  {Dawson}, {Frye}, {Hoag}, {Huang}, {Johnson}, {Jones}, {Kikuchihara}, {Lam},
  {Livermore}, {Lovisari}, {Mainali}, {Oesch}, {Ogaz}, {Ouchi}, {Past},
  {Paterno-Mahler}, {Peterson}, {Ryan}, {Salmon}, {Sendra-Server}, {Stark},
  {Strait}, {Toft}, \& {Vulcani}}]{Acebron2019}
{Acebron}, A., {Alon}, M., {Zitrin}, A., {et~al.} 2019,
  \bibinfo{title}{{RELICS: High-resolution Constraints on the Inner Mass
  Distribution of the z = 0.83 Merging Cluster RXJ0152.7-1357 from Strong
  Lensing},} The Astrophysical Journal, 874, 132,
  \dodoi{10.3847/1538-4357/ab0adf}

\bibitem[{M. Akhshik {et~al.}(2023)Akhshik, Whitaker, Leja, Richard, Spilker,
  Song, Brammer, Bezanson, Ebeling, Gallazzi, Mahler, Mowla, Nelson, Pacifici,
  Sharon, Toft, Williams, Wright, \& Zabl}]{akhshik_requiem-2d_2023}
Akhshik, M., Whitaker, K.~E., Leja, J., {et~al.} 2023,
  \bibinfo{title}{{REQUIEM}-{2D}: {A} {Diversity} of {Formation} {Pathways} in
  a {Sample} of {Spatially} {Resolved} {Massive} {Quiescent} {Galaxies} at z
  $\sim$ 2,} The Astrophysical Journal, 943, 179,
  \dodoi{10.3847/1538-4357/aca677}

\bibitem[{D.~M. Alexander {et~al.}(2025)Alexander, Hickox, Aird, Combes, Costa,
  Habouzit, Harrison, Leng, Morabito, Uckelman, \&
  Vickers}]{alexander_what_2025}
Alexander, D.~M., Hickox, R.~C., Aird, J., {et~al.} 2025, What drives the
  growth of black holes: a decade of progress, arXiv,
  \dodoi{10.48550/arXiv.2506.19166}

\bibitem[{J.~F.~V. {Allingham} {et~al.}(2026){Allingham}, {Zitrin}, {Kokorev},
  {Yanagisawa}, {Diego}, {Furtak}, {Asada}, {Coe}, {Coulter}, {Fujimoto},
  {Larison}, {Oguri}, {Pierel}, {Sun}, {Bradac}, {Dayal}, {Lopes}, {Meena},
  {Pascale}, {Akins}, {Bauer}, {Bradley}, {Brammer}, {Chisholm}, {Desprez},
  {Fei}, {Ferguson}, {Finkelstein}, {Frye}, {Golubchik}, {Inayoshi},
  {Koekemoer}, {Lucas}, {Magdis}, {Martis}, {Pan}, {Richard}, {Ricotti},
  {Rihtarsic}, {Robbins}, {Sheu}, {Welch}, {Willott}, \&
  {Windhorst}}]{Allingham26a}
{Allingham}, J. F.~V., {Zitrin}, A., {Kokorev}, V., {et~al.} 2026,
  \bibinfo{title}{{VENUS: Strong-lensing model of MACS J1931.8-2635 --
  revealing the farthest multiply imaged supernova},} arXiv e-prints,
  arXiv:2602.14074, \dodoi{10.48550/arXiv.2602.14074}

\bibitem[{J.~F.~V. Allingham {et~al.}(2026)Allingham, Furtak, Golubchik, Meena,
  Zitrin, Chisholm, Fujimoto, Kokorev, {Abdurro'uf}, Bauer, Bradač, Brammer,
  Coe, Diego, Ferguson, Finkelstein, Hsiao, Jiménez-Teja, Palencia,
  Rihtaršič, Willott, Windhorst, Akins, Amorín, Bradley, Cheng, Conselice,
  Coulter, Dayal, Desprez, Dessauges-Zavadsky, Faisst, Fei, Frye, Harikane,
  Kartaltepe, Kiyota, Larison, Lopes, Lucas, Magdis, Martis, Ouchi, Pan,
  Papovich, Pascale, Pierel, Resseguier, Richard, Ricotti, Robbins, Sheu, Sun,
  Vecchi, Welch, \& Yanagisawa}]{allinghamVENUSCensusStrong2026}
Allingham, J. F.~V., Furtak, L.~J., Golubchik, M., {et~al.} 2026, {VENUS}: {A}
  {Census} of {Strong} {Lensing} in a 60-{Cluster} {JWST} {Survey}, arXiv,
  \dodoi{10.48550/arXiv.2611.33142}

\bibitem[{J. Antwi-Danso {et~al.}(2023)Antwi-Danso, Papovich, Leja, Marchesini,
  Marsan, Martis, Labb{\'e}, Muzzin, Glazebrook, Straatman, \&
  Tran}]{antwi-danso_beyond_2023}
Antwi-Danso, J., Papovich, C., Leja, J., {et~al.} 2023, \bibinfo{title}{Beyond
  {UVJ}: {Color} {Selection} of {Galaxies} in the {JWST} {Era},} The
  Astrophysical Journal, 943, 166, \dodoi{10.3847/1538-4357/aca294}

\bibitem[{J. Antwi-Danso {et~al.}(2025)Antwi-Danso, Papovich, Esdaile,
  Nanayakkara, Glazebrook, Hutchison, Whitaker, Marsan, Diaz, Marchesini,
  Muzzin, Tran, Setton, Kaushal, Speagle, \& Cole}]{antwi-danso_feniks_2025}
Antwi-Danso, J., Papovich, C., Esdaile, J., {et~al.} 2025, \bibinfo{title}{The
  {FENIKS} {Survey}: {Spectroscopic} {Confirmation} of {Massive} {Quiescent}
  {Galaxies} at z \textasciitilde{} 3--5,} The Astrophysical Journal, 978, 90,
  \dodoi{10.3847/1538-4357/ad8b30}

\bibitem[{ {Astropy Collaboration} {et~al.}(2022){Astropy Collaboration},
  Price-Whelan, Lim, Earl, Starkman,
  {et~al.}}]{astropy_collaboration_astropy_2022}
{Astropy Collaboration}, Price-Whelan, A.~M., Lim, P.~L., {et~al.} 2022,
  \bibinfo{title}{The {Astropy} {Project}: {Sustaining} and {Growing} a
  {Community}-oriented {Open}-source {Project} and the {Latest} {Major}
  {Release} (v5.0) of the {Core} {Package},} The Astrophysical Journal, 935,
  167, \dodoi{10.3847/1538-4357/ac7c74}

\bibitem[{H. Atek {et~al.}(2025)Atek, Chisholm, Kokorev, Endsley, Pan, Furtak,
  Chemerynska, Richard, Claeyssens, Oesch, Fujimoto, Naidu, Korber, Schaerer,
  Blaizot, Rosdahl, Adamo, Asada, Basu, Beauchesne, Berg, Bezanson, Bouwens,
  Brammer, Dessauges-Zavadsky, Ellien, Ezziati, Fei, Goovaerts, Heurtier,
  Hsiao, Jecmen, Khullar, Kneib, Labb{\'e}, Leclercq, Marques-Chaves, Mason,
  McQuinn, Muñoz, Natarajan, Saldana-Lopez, Stephenson, Trebitsch, Volonteri,
  Weibel, \& Zitrin}]{atek_jwsts_2025}
Atek, H., Chisholm, J., Kokorev, V., {et~al.} 2025, {JWST}'s {GLIMPSE}: an
  overview of the deepest probe of early galaxy formation and cosmic
  reionization, arXiv, \dodoi{10.48550/arXiv.2511.07542}

\bibitem[{J.~F.~W. {Baggen} {et~al.}(2024){Baggen}, {van Dokkum}, {Brammer},
  {de Graaff}, {Franx}, {Greene}, {Labb{\'e}}, {Leja}, {Maseda}, {Nelson},
  {Rix}, {Wang}, \& {Weibel}}]{baggen_small_2024}
{Baggen}, J. F.~W., {van Dokkum}, P., {Brammer}, G., {et~al.} 2024,
  \bibinfo{title}{{The Small Sizes and High Implied Densities of ``Little Red
  Dots'' with Balmer Breaks Could Explain Their Broad Emission Lines without an
  Active Galactic Nucleus},} The Astrophysical Journal Letters, 977, L13,
  \dodoi{10.3847/2041-8213/ad90b8}

\bibitem[{W.~M. Baker {et~al.}(2025)Baker, Lim, D'Eugenio, Maiolino, Ji,
  Arribas, Bunker, Carniani, Charlot, de~Graaff, Hainline, Looser, Lyu,
  Rinaldi, Robertson, Schaller, Schaye, Scholtz, Übler, Williams, Willmer,
  Willott, \& Zhu}]{baker_abundance_2025}
Baker, W.~M., Lim, S., D'Eugenio, F., {et~al.} 2025, \bibinfo{title}{The
  abundance and nature of high-redshift quiescent galaxies from {JADES}
  spectroscopy and the {FLAMINGO} simulations,} Monthly Notices of the Royal
  Astronomical Society, 539, 557, \dodoi{10.1093/mnras/staf475}

\bibitem[{G. Barro {et~al.}(2017)Barro, Faber, Koo, Dekel, Fang,
  {et~al.}}]{barro_structural_2017}
Barro, G., Faber, S.~M., Koo, D.~C., {et~al.} 2017, \bibinfo{title}{Structural
  and {Star}-forming {Relations} since $z \sim 3$: {Connecting} {Compact}
  {Star}-forming and {Quiescent} {Galaxies},} The Astrophysical Journal, 840,
  47, \dodoi{10.3847/1538-4357/aa6b05}

\bibitem[{E. Bertin \& S. Arnouts(1996)Bertin \&
  Arnouts}]{bertin_sextractor_1996}
Bertin, E., \& Arnouts, S. 1996, \bibinfo{title}{{SExtractor}: {Software} for
  source extraction.,} Astronomy and Astrophysics Supplement Series, 117, 393,
  \dodoi{10.1051/aas:1996164}

\bibitem[{M.~R. Blanton \& S. Roweis(2007)Blanton \&
  Roweis}]{blanton_k-corrections_2007}
Blanton, M.~R., \& Roweis, S. 2007, \bibinfo{title}{K-{Corrections} and
  {Filter} {Transformations} in the {Ultraviolet}, {Optical}, and
  {Near}-{Infrared},} The Astronomical Journal, 133, 734,
  \dodoi{10.1086/510127}

\bibitem[{G. Brammer(2023)Brammer}]{grizli}
Brammer, G. 2023, grizli, 1.9.11 Zenodo, \dodoi{10.5281/zenodo.8370018}

\bibitem[{G.~B. Brammer {et~al.}(2008)Brammer, van Dokkum, \&
  Coppi}]{brammer_eazy_2008}
Brammer, G.~B., van Dokkum, P.~G., \& Coppi, P. 2008, \bibinfo{title}{{EAZY}:
  {A} {Fast}, {Public} {Photometric} {Redshift} {Code},} The Astrophysical
  Journal, 686, 1503, \dodoi{10.1086/591786}

\bibitem[{M. Cappellari \& Y. Copin(2003)Cappellari \&
  Copin}]{cappellari_adaptive_2003}
Cappellari, M., \& Copin, Y. 2003, \bibinfo{title}{Adaptive spatial binning of
  integral-field spectroscopic data using {Voronoi} tessellations,} Monthly
  Notices of the Royal Astronomical Society, 342, 345,
  \dodoi{10.1046/j.1365-8711.2003.06541.x}

\bibitem[{A.~C. Carnall {et~al.}(2023{\natexlab{a}})Carnall, McLure, Dunlop,
  McLeod, Wild, Cullen, Magee, Begley, Cimatti, Donnan, Hamadouche, Jewell, \&
  Walker}]{carnall_massive_2023}
Carnall, A.~C., McLure, R.~J., Dunlop, J.~S., {et~al.} 2023{\natexlab{a}},
  \bibinfo{title}{A massive quiescent galaxy at redshift 4.658,} Nature, 619,
  716, \dodoi{10.1038/s41586-023-06158-6}

\bibitem[{A.~C. Carnall {et~al.}(2023{\natexlab{b}})Carnall, McLeod, McLure,
  Dunlop, Begley, Cullen, Donnan, Hamadouche, Jewell, Jones, Pollock, \&
  Wild}]{carnallSurprisingAbundanceMassive2023}
Carnall, A.~C., McLeod, D.~J., McLure, R.~J., {et~al.} 2023{\natexlab{b}},
  \bibinfo{title}{A surprising abundance of massive quiescent galaxies at 3
  {\textless} z {\textless} 5 in the first data from {JWST} {CEERS},} Monthly
  Notices of the Royal Astronomical Society, 520, 3974,
  \dodoi{10.1093/mnras/stad369}

\bibitem[{A.~C. Carnall {et~al.}(2024)Carnall, Cullen, McLure, McLeod, Begley,
  Donnan, Dunlop, Shapley, Rowlands, Almaini, Arellano-Córdova, Barrufet,
  Cimatti, Ellis, Grogin, Hamadouche, Illingworth, Koekemoer, Leung, Lovell,
  P{\'e}rez-Gonz{\'a}lez, Santini, Stanton, \& Wild}]{carnall_jwst_2024}
Carnall, A.~C., Cullen, F., McLure, R.~J., {et~al.} 2024, \bibinfo{title}{The
  {JWST} {EXCELS} survey: too much, too young, too fast? {Ultra}-massive
  quiescent galaxies at 3 {\textless} z {\textless} 5,} Monthly Notices of the
  Royal Astronomical Society, 534, 325, \dodoi{10.1093/mnras/stae2092}

\bibitem[{C. {Cerny} {et~al.}(2026){Cerny}, {Mahler}, {Sharon}, {Jauzac},
  {Khullar}, {Beauchesne}, {Diego}, {Lagattuta}, {Limousin}, {Patel},
  {Richard}, {Cornil-Ba{\"\i}otto}, {Gladders}, {Werner}, {Doppel}, {Floyd},
  {Gonzalez}, {Massey}, {Montes}, {Bayliss}, {Bleem}, {Canning}, {Edge},
  {McDonald}, {Natarajan}, {Stark}, \& {Gassis}}]{Cerny2026}
{Cerny}, C., {Mahler}, G., {Sharon}, K., {et~al.} 2026, \bibinfo{title}{{Strong
  LensIng and Cluster Evolution (SLICE) with JWST: Early Results, Lens Models,
  and High-redshift Detections},} The Astrophysical Journal, 1001, 60,
  \dodoi{10.3847/1538-4357/ae41b0}

\bibitem[{G. Chabrier(2003)Chabrier}]{chabrier_galactic_2003}
Chabrier, G. 2003, \bibinfo{title}{Galactic {Stellar} and {Substellar}
  {Initial} {Mass} {Function},} Publications of the Astronomical Society of the
  Pacific, 115, 763, \dodoi{10.1086/376392}

\bibitem[{{\'A}. {Chandro-G{\'o}mez} {et~al.}(2025){Chandro-G{\'o}mez},
  {Lagos}, {Power}, {Baker}, {Ben{\'\i}tez-Llambay}, {Chaikin}, {Chittenden},
  {Correa}, {Frenk}, {Hu{\v{s}}ko}, {McGibbon}, {Nanayakkara}, {Ploeckinger},
  {Richings}, {Schaller}, {Schaye}, \& {Trayford}}]{chandro2025}
{Chandro-G{\'o}mez}, {\'A}., {Lagos}, C. d.~P., {Power}, C., {et~al.} 2025,
  \bibinfo{title}{{Unveiling the population of massive quenched galaxies at
  $z\ge2$ in the COLIBRE simulations -- I. Galaxy demographics},} arXiv
  e-prints, arXiv:2512.16208, \dodoi{10.48550/arXiv.2512.16208}

\bibitem[{{\'A}. Chandro-G{\'o}mez {et~al.}(2026)Chandro-G{\'o}mez, Lagos,
  Power, Baker, Benítez-Llambay, Chaikin, Chittenden, Correa, Frenk, Huško,
  Ito, McGibbon, Nanayakkara, Ploeckinger, Richings, Schaller, Schaye,
  Trayford, \& Valentino}]{chandro-gomez_unveiling_2026}
Chandro-G{\'o}mez, {\'A}., Lagos, C. d.~P., Power, C., {et~al.} 2026, Unveiling
  the population of massive quenched galaxies at $z\ge2$ in the {COLIBRE}
  simulations -- {II}. {The} role of {AGN} feedback and environment on their
  emergence, arXiv, \dodoi{10.48550/arXiv.2605.31052}

\bibitem[{C.~M. Cheng {et~al.}(2024)Cheng, Kriek, Beverage, van der Wel,
  Bezanson, D’Eugenio, Franx, Mancera Piña, Nersesian, Slob, Suess,
  van Dokkum, Wu, Gallazzi, \& Zibetti}]{cheng_age_2024}
Cheng, C.~M., Kriek, M., Beverage, A.~G., {et~al.} 2024, \bibinfo{title}{Age
  and metal gradients in massive quiescent galaxies at 0.6 $\lesssim$ z
  $\lesssim$ 1.0: implications for quenching and assembly histories,} Monthly
  Notices of the Royal Astronomical Society, 532, 3604,
  \dodoi{10.1093/mnras/stae1739}

\bibitem[{J. Choi {et~al.}(2016)Choi, Dotter, Conroy, Cantiello, Paxton, \&
  Johnson}]{choi_mesa_2016}
Choi, J., Dotter, A., Conroy, C., {et~al.} 2016, \bibinfo{title}{{MESA}
  {ISOCHRONES} {AND} {STELLAR} {TRACKS} ({MIST}). {I}. {SOLAR}-{SCALED}
  {MODELS},} The Astrophysical Journal, 823, 102,
  \dodoi{10.3847/0004-637X/823/2/102}

\bibitem[{D. Coe {et~al.}(2019)Coe, Salmon, Bradač, Bradley, Sharon, Zitrin,
  Acebron, Cerny, Cibirka, Strait, Paterno-Mahler, Mahler, Avila, Ogaz, Huang,
  Pelliccia, Stark, Mainali, Oesch, Trenti, Carrasco, Dawson, Rodney, Strolger,
  Riess, Jones, Frye, Czakon, Umetsu, Vulcani, Graur, Jha, Graham, Molino,
  Nonino, Hjorth, Selsing, Christensen, Kikuchihara, Ouchi, Oguri, Welch,
  Lemaux, Andrade-Santos, Hoag, Johnson, Peterson, Past, Fox, Agulli,
  Livermore, Ryan, Lam, Sendra-Server, Toft, Lovisari, \& Su}]{coe_relics_2019}
Coe, D., Salmon, B., Bradač, M., {et~al.} 2019, \bibinfo{title}{{RELICS}:
  {Reionization} {Lensing} {Cluster} {Survey},} The Astrophysical Journal, 884,
  85, \dodoi{10.3847/1538-4357/ab412b}

\bibitem[{C. Conroy {et~al.}(2009)Conroy, Gunn, \&
  White}]{conroy_propagation_2009}
Conroy, C., Gunn, J.~E., \& White, M. 2009, \bibinfo{title}{The {Propagation}
  of {Uncertainties} in {Stellar} {Population} {Synthesis} {Modeling}. {I}.
  {The} {Relevance} of {Uncertain} {Aspects} of {Stellar} {Evolution} and the
  {Initial} {Mass} {Function} to the {Derived} {Physical} {Properties} of
  {Galaxies},} The Astrophysical Journal, 699, 486,
  \dodoi{10.1088/0004-637X/699/1/486}

\bibitem[{C. {Conroy} {et~al.}(2009){Conroy}, {Gunn}, \& {White}}]{fsps2009}
{Conroy}, C., {Gunn}, J.~E., \& {White}, M. 2009, \bibinfo{title}{{The
  Propagation of Uncertainties in Stellar Population Synthesis Modeling. I. The
  Relevance of Uncertain Aspects of Stellar Evolution and the Initial Mass
  Function to the Derived Physical Properties of Galaxies},} The Astrophysical
  Journal, 699, 486, \dodoi{10.1088/0004-637X/699/1/486}

\bibitem[{D.~J. Croton {et~al.}(2006)Croton, Springel, White, De~Lucia, Frenk,
  Gao, Jenkins, Kauffmann, Navarro, \& Yoshida}]{crotonManyLivesActive2006}
Croton, D.~J., Springel, V., White, S. D.~M., {et~al.} 2006,
  \bibinfo{title}{The many lives of active galactic nuclei: cooling flows,
  black holes and the luminosities and colours of galaxies,} Monthly Notices of
  the Royal Astronomical Society, 365, 11,
  \dodoi{10.1111/j.1365-2966.2005.09675.x}

\bibitem[{S.~E. Cutler {et~al.}(2026)Cutler, Robbins, Marchesini, Suess,
  Muzzin, Brammer, Asada, Martis, Alberts, Antwi-Danso, Cloonan, Labb{\'e},
  Miller, Mitsuhashi, Pope, Sajina, Sarrouh, Sharma, Stefanon, Vidal, Willot,
  Bezanson, Bradač, Cooper, Feldmann, Forrest, Glazebrook, Greene, La~Torre,
  Lin, Maseda, McConachie, Nananyakkara, Noirot, Pan, Patel, Pratt, Sawicki,
  Setton, Weaver, van~der Wel, Whitaker, Zhang, \&
  Zaidi}]{cutler_unbreaking_2026}
Cutler, S.~E., Robbins, L., Marchesini, D., {et~al.} 2026, Unbreaking the
  {Universe}: {MINERVA} {Measurements} of {Color} {Gradients} in {Massive}
  {Quiescent} {Galaxies} {Can} {Help} {Ease} {Too}-{Early} {Star} {Formation}
  {Tensions}, arXiv, \dodoi{10.48550/arXiv.2606.02698}

\bibitem[{A. de~Graaff {et~al.}(2025)de~Graaff, Setton, Brammer, Cutler, Suess,
  Labb{\'e}, Leja, Weibel, Maseda, Whitaker, Bezanson, Boogaard, Cleri,
  De~Lucia, Franx, Greene, Hirschmann, Matthee, McConachie, Naidu, Oesch,
  Price, Rix, Valentino, Wang, \& Williams}]{de_graaff_efficient_2025}
de~Graaff, A., Setton, D.~J., Brammer, G., {et~al.} 2025,
  \bibinfo{title}{Efficient formation of a massive quiescent galaxy at redshift
  4.9,} Nature Astronomy, 9, 280, \dodoi{10.1038/s41550-024-02424-3}

\bibitem[{R. {Demarco} {et~al.}(2005){Demarco}, {Rosati}, {Lidman}, {Homeier},
  {Scannapieco}, {Ben{\'\i}tez}, {Mainieri}, {Nonino}, {Girardi}, {Stanford},
  {Tozzi}, {Borgani}, {Silk}, {Squires}, \& {Broadhurst}}]{Demarco2005}
{Demarco}, R., {Rosati}, P., {Lidman}, C., {et~al.} 2005, \bibinfo{title}{{A
  VLT spectroscopic survey of RX J0152.7-1357, a forming cluster of galaxies at
  z = 0.837},} Astronomy and Astrophysics, 432, 381,
  \dodoi{10.1051/0004-6361:20041931}

\bibitem[{F. D'Eugenio {et~al.}(2025)D'Eugenio, Helton, Hainline, Sun,
  Maiolino, P{\'e}rez-Gonz{\'a}lez, Juodžbalis, Arribas, Bunker, Carniani,
  Curtis-Lake, Egami, Eisenstein, Johnson, Robertson, Tacchella, Willmer,
  Willott, Baker, Danhaive, Duan, Fudamoto, Jones, Lin, Liu, Perna, Pusk{\'a}s,
  Rinaldi, Scholtz, Sun, Trussler, Übler, Venturi, Williams, \&
  Zhu}]{deugenioJADESSAPPHIRESGalaxy2025}
D'Eugenio, F., Helton, J.~M., Hainline, K., {et~al.} 2025,
  \bibinfo{title}{{JADES} and {SAPPHIRES}: galaxy metamorphosis amidst a huge,
  luminous emission-line region,} Monthly Notices of the Royal Astronomical
  Society, 542, 960, \dodoi{10.1093/mnras/staf1138}

\bibitem[{F. D'Eugenio {et~al.}(2026)D'Eugenio, Brazzini, Maiolino, Bertola,
  Carniani, Ji, Juodžbalis, Liu, Park, Parlanti, Pascalau, Pezzulli, Scholtz,
  Tacchella, \& Zibetti}]{deugenioDustyQuenchingcandidateAGN2026}
D'Eugenio, F., Brazzini, M., Maiolino, R., {et~al.} 2026, A {Dusty}
  {Quenching}-candidate {AGN} host at z=5.7: massive quiescent galaxies may
  quench already during the dust-obscured phase, arXiv

\bibitem[{A. Dotter(2016)Dotter}]{dotter_mesa_2016}
Dotter, A. 2016, \bibinfo{title}{{MESA} {ISOCHRONES} {AND} {STELLAR} {TRACKS}
  ({MIST}) 0: {METHODS} {FOR} {THE} {CONSTRUCTION} {OF} {STELLAR}
  {ISOCHRONES},} The Astrophysical Journal Supplement Series, 222, 8,
  \dodoi{10.3847/0067-0049/222/1/8}

\bibitem[{K.~J. Duncan {et~al.}(2023)Duncan, Windhorst, Koekemoer,
  R{\"o}ttgering, Cohen, Jansen, Summers, Tompkins, Hutchison, Conselice,
  Driver, Yan, Adams, Cheng, Coe, Diego, Dole, Frye, Gim, Grogin, Holwerda,
  Lim, Marshall, Nonino, Pirzkal, Robotham, Ryan, \&
  Willmer}]{duncan_jwsts_2023}
Duncan, K.~J., Windhorst, R.~A., Koekemoer, A.~M., {et~al.} 2023,
  \bibinfo{title}{{JWST}'s {PEARLS}: {TN} {J1338}-1942 - {I}. {Extreme}
  jet-triggered star formation in a z = 4.11 luminous radio galaxy,} Monthly
  Notices of the Royal Astronomical Society, 522, 4548,
  \dodoi{10.1093/mnras/stad1267}

\bibitem[{A.~H. Edward {et~al.}(2026)Edward, Antwi-Danso, Muzzin, Forrest,
  McConachie, Beverage, Chang, Cooper, Gomez, Hamadouche, Henry, Lei,
  Marchesini, Noble, Urbano~Stawinski, Wilson, \& Wisz}]{edward_magaz3ne_2026}
Edward, A.~H., Antwi-Danso, J., Muzzin, A., {et~al.} 2026, {MAGAZ3NE}:
  {Spatially} {Resolved} {Ages} and {Chemical} {Abundances} of
  {Ultra}-{Massive} {Quiescent} {Galaxies} at z $\sim$ 3.5 using
  {JWST}/{NIRSpec} {IFU}, arXiv, \dodoi{10.48550/arXiv.2605.27555}

\bibitem[{S.~M. {Faber} \& R.~E. {Jackson}(1976){Faber} \&
  {Jackson}}]{FaberJackson}
{Faber}, S.~M., \& {Jackson}, R.~E. 1976, \bibinfo{title}{{Velocity dispersions
  and mass-to-light ratios for elliptical galaxies.},} The Astrophysical
  Journal, 204, 668, \dodoi{10.1086/154215}

\bibitem[{A.~C. Fabian(2012)Fabian}]{fabianObservationalEvidenceActive2012}
Fabian, A.~C. 2012, \bibinfo{title}{Observational {Evidence} of {Active}
  {Galactic} {Nuclei} {Feedback},} Annual Review of Astronomy and Astrophysics,
  50, 455, \dodoi{10.1146/annurev-astro-081811-125521}

\bibitem[{A.~L. Faisst \& T. Morishita(2024)Faisst \&
  Morishita}]{faisstDeadAliveHow2024}
Faisst, A.~L., \& Morishita, T. 2024, \bibinfo{title}{Dead or {Alive}? {How}
  {Bursty} {Star} {Formation} and {Patchy} {Dust} {Can} {Cause} {Temporary}
  {Quiescence} in {High}-redshift {Galaxies},} The Astrophysical Journal, 971,
  47, \dodoi{10.3847/1538-4357/ad58e2}

\bibitem[{D. {Foreman-Mackey} {et~al.}(2014){Foreman-Mackey}, {Sick}, \&
  {Johnson}}]{pythonfsps2014}
{Foreman-Mackey}, D., {Sick}, J., \& {Johnson}, B. 2014, {python-fsps: Python
  bindings to FSPS (v0.1.1)}, v0.1.1 Zenodo, \dodoi{10.5281/zenodo.12157}

\bibitem[{B. Forrest {et~al.}(2026)Forrest, Muzzin, Marchesini, Pan, Ozden,
  Antwi-Danso, Chang, Cooper, Edward, Gomez, Kimmig, Lemaux, McConachie, Noble,
  Remus, Urbano~Stawinski, Wilson, \&
  Wisz}]{forrestMassiveEvolvedSlowrotating2026}
Forrest, B., Muzzin, A., Marchesini, D., {et~al.} 2026, \bibinfo{title}{A
  massive and evolved slow-rotating galaxy in the early {Universe},} Nature
  Astronomy, \dodoi{10.1038/s41550-026-02855-0}

\bibitem[{L.~J. {Furtak} {et~al.}(2023){Furtak}, {Zitrin}, {Weaver}, {Atek},
  {Bezanson}, {Labb{\'e}}, {Whitaker}, {Leja}, {Price}, {Brammer}, {Wang},
  {Marchesini}, {Pan}, {Dayal}, {van Dokkum}, {Feldmann}, {Fujimoto}, {Franx},
  {Khullar}, {Nelson}, \& {Mowla}}]{Furtak2022UNCOVER}
{Furtak}, L.~J., {Zitrin}, A., {Weaver}, J.~R., {et~al.} 2023,
  \bibinfo{title}{{UNCOVERing the extended strong lensing structures of Abell
  2744 with the deepest JWST imaging},} Monthly Notices of the Royal
  Astronomical Society, 523, 4568, \dodoi{10.1093/mnras/stad1627}

\bibitem[{L.~J. {Furtak} {et~al.}(2024){Furtak}, {Labb{\'e}}, {Zitrin},
  {Greene}, {Dayal}, {Chemerynska}, {Kokorev}, {Miller}, {Goulding}, {de
  Graaff}, {Bezanson}, {Brammer}, {Cutler}, {Leja}, {Pan}, {Price}, {Wang},
  {Weaver}, {Whitaker}, {Atek}, {Bogd{\'a}n}, {Charlot}, {Curtis-Lake}, {van
  Dokkum}, {Endsley}, {Feldmann}, {Fudamoto}, {Fujimoto}, {Glazebrook},
  {Juneau}, {Marchesini}, {Maseda}, {Nelson}, {Oesch}, {Plat}, {Setton},
  {Stark}, \& {Williams}}]{furtak_high_2024}
{Furtak}, L.~J., {Labb{\'e}}, I., {Zitrin}, A., {et~al.} 2024,
  \bibinfo{title}{{A high black-hole-to-host mass ratio in a lensed AGN in the
  early Universe},} Nature, 628, 57, \dodoi{10.1038/s41586-024-07184-8}

\bibitem[{K. Glazebrook {et~al.}(2024)Glazebrook, Nanayakkara, Schreiber,
  Lagos, Kawinwanichakij, Jacobs, Chittenden, Brammer, Kacprzak, Labbe,
  Marchesini, Marsan, Oesch, Papovich, Remus, Tran, Esdaile, \&
  Chandro-Gomez}]{glazebrook_massive_2024}
Glazebrook, K., Nanayakkara, T., Schreiber, C., {et~al.} 2024,
  \bibinfo{title}{A massive galaxy that formed its stars at z $\simeq$ 11,}
  Nature, 628, 277, \dodoi{10.1038/s41586-024-07191-9}

\bibitem[{M. {Golubchik} {et~al.}(2025){Golubchik}, {Furtak}, {Allingham},
  {Zitrin}, {Akins}, {Kokorev}, {Fujimoto}, {Abdurro'uf}, {Amor{\'\i}n},
  {Bauer}, {Bezanson}, {Brada{\v{c}}}, {Bradley}, {Brammer}, {Chisholm}, {Coe},
  {Conselice}, {Dayal}, {Dessauges-Zavadsky}, {Diego}, {Faisst}, {Fei},
  {Ferguson}, {Finkelstein}, {Frye}, {Gonz{\'a}lez-Otero}, {Greene},
  {Harikane}, {Hsiao}, {Inayoshi}, {Jim{\'e}nez-Teja}, {Knudsen}, {Koekemoer},
  {Labb{\'e}}, {Lucas}, {Magdis}, {Matthee}, {Messa}, {Naidu}, {Nakane},
  {Noirot}, {Pan}, {Papovich}, {Richard}, {Ricotti}, {Robbins}, {Stark}, {Sun},
  {Treu}, {Tripodi}, {Vanzella}, {Willott}, \&
  {Windhorst}}]{Golubchik2025_LRD_VENUS}
{Golubchik}, M., {Furtak}, L.~J., {Allingham}, J. F.~V., {et~al.} 2025,
  \bibinfo{title}{{VENUS: When Red meets Blue -- A multiply imaged Little Red
  Dot with an apparent blue companion behind the galaxy cluster Abell 383},}
  arXiv e-prints, arXiv:2512.02117.
\newblock \doarXiv{2512.02117}

\bibitem[{C. {Grillo} {et~al.}(2008){Grillo}, {Lombardi}, {Rosati}, {Bertin},
  {Gobat}, {Demarco}, {Lidman}, {Motta}, \& {Nonino}}]{Grillo2008}
{Grillo}, C., {Lombardi}, M., {Rosati}, P., {et~al.} 2008, \bibinfo{title}{{A
  twelve-image gravitational lens system in the z $\simeq$ 0.84 cluster Cl
  J0152.7-1357},} Astronomy and Astrophysics, 486, 45,
  \dodoi{10.1051/0004-6361:200809434}

\bibitem[{R.~C. Hickox \& D.~M. Alexander(2018)Hickox \&
  Alexander}]{hickox_obscured_2018}
Hickox, R.~C., \& Alexander, D.~M. 2018, \bibinfo{title}{Obscured {Active}
  {Galactic} {Nuclei},} Annual Review of Astronomy and Astrophysics, 56, 625,
  \dodoi{10.1146/annurev-astro-081817-051803}

\bibitem[{G. Hinshaw {et~al.}(2013)Hinshaw, Larson, Komatsu, Spergel, Bennett,
  {et~al.}}]{hinshaw_nine-year_2013}
Hinshaw, G., Larson, D., Komatsu, E., {et~al.} 2013, \bibinfo{title}{Nine-year
  {Wilkinson} {Microwave} {Anisotropy} {Probe} ({WMAP}) {Observations}:
  {Cosmological} {Parameter} {Results},} The Astrophysical Journal Supplement
  Series, 208, 19, \dodoi{10.1088/0067-0049/208/2/19}

\bibitem[{P.~F. Hopkins {et~al.}(2008)Hopkins, Hernquist, Cox, \&
  Kereš}]{hopkinsCosmologicalFrameworkCoEvolution2008}
Hopkins, P.~F., Hernquist, L., Cox, T.~J., \& Kereš, D. 2008,
  \bibinfo{title}{A {Cosmological} {Framework} for the {Co}-{Evolution} of
  {Quasars}, {Supermassive} {Black} {Holes}, and {Elliptical} {Galaxies}. {I}.
  {Galaxy} {Mergers} and {Quasar} {Activity},} The Astrophysical Journal
  Supplement Series, 175, 356, \dodoi{10.1086/524362}

\bibitem[{P.~F. Hopkins {et~al.}(2006)Hopkins, Somerville, Hernquist, Cox,
  Robertson, \& Li}]{hopkinsRelationQuasarMerging2006}
Hopkins, P.~F., Somerville, R.~S., Hernquist, L., {et~al.} 2006,
  \bibinfo{title}{The {Relation} between {Quasar} and {Merging} {Galaxy}
  {Luminosity} {Functions} and the {Merger}-driven {Star} {Formation} {History}
  of the {Universe},} The Astrophysical Journal, 652, 864,
  \dodoi{10.1086/508503}

\bibitem[{T.~A. Hutchison {et~al.}(2025)Hutchison, Khullar, Rigby, Welch,
  Florian, Sharon, Sierra, Sarmiento, Mahler, Cleri, Bezanson, Gladders,
  Bayliss, Karp, Berry, Ross, Rivera-Thorsen, Choe, Dahle, Chisholm, Lambrides,
  Larson, Olivier, Owens, \& Solhaug}]{hutchisonJWSTWazArc2025}
Hutchison, T.~A., Khullar, G., Rigby, J.~R., {et~al.} 2025, {JWST} \& the {Waz}
  {Arc} {I}: {Spatially} {Resolving} the {Physical} {Conditions} within a
  {Post}-{Starburst} {Galaxy} at {Redshift} 5 with {NIRSpec} {IFS}, arXiv,
  \dodoi{10.48550/arXiv.2512.02000}

\bibitem[{K. Ito {et~al.}(2025)Ito, Valentino, Brammer, Hamadouche, Whitaker,
  Kokorev, Zhu, Kakimoto, Wu, Antwi-Danso, Baker, Ceverino, Faisst, Farcy,
  Fujimoto, Gallazzi, Gillman, Gottumukkala, Heintz, Hirschmann, Jespersen,
  Kubo, Lee, Magdis, Onodera, Shimakawa, Tanaka, Toft, \&
  Weaver}]{ito_deepdive_2025}
Ito, K., Valentino, F., Brammer, G., {et~al.} 2025, {DeepDive}: {A} deep dive
  into the physics of the first massive quiescent galaxies in the {Universe},
  arXiv, \dodoi{10.48550/arXiv.2506.22642}

\bibitem[{M. Jafariyazani {et~al.}(2025)Jafariyazani, Newman, Mobasher, Belli,
  Ellis, \& Faisst}]{jafariyazaniChemicalAbundancesEarly2025}
Jafariyazani, M., Newman, A.~B., Mobasher, B., {et~al.} 2025,
  \bibinfo{title}{Chemical {Abundances} of {Early} {Quiescent} {Galaxies}:
  {New} {Observations} and {Modeling} {Impacts},} The Astrophysical Journal,
  986, 148, \dodoi{10.3847/1538-4357/addbdc}

\bibitem[{M. Jafariyazani {et~al.}(2020)Jafariyazani, Newman, Mobasher, Belli,
  Ellis, \& Patel}]{jafariyazani_resolved_2020}
Jafariyazani, M., Newman, A.~B., Mobasher, B., {et~al.} 2020,
  \bibinfo{title}{Resolved {Multi}-element {Stellar} {Chemical} {Abundances} in
  the {Brightest} {Quiescent} {Galaxy} at z $\sim$ 2,} The Astrophysical
  Journal Letters, 897, L42, \dodoi{10.3847/2041-8213/aba11c}

\bibitem[{Z. Ji \& M. Giavalisco(2023)Ji \&
  Giavalisco}]{ji_reconstructing_2023}
Ji, Z., \& Giavalisco, M. 2023, \bibinfo{title}{Reconstructing the {Assembly}
  of {Massive} {Galaxies}. {II}. {Galaxies} {Develop} {Massive} and {Dense}
  {Stellar} {Cores} as {They} {Evolve} and {Head} toward {Quiescence} at
  {Cosmic} {Noon},} The Astrophysical Journal, 943, 54,
  \dodoi{10.3847/1538-4357/aca807}

\bibitem[{B.~D. Johnson {et~al.}(2021)Johnson, Leja, Conroy, \&
  Speagle}]{johnson_stellar_2021}
Johnson, B.~D., Leja, J., Conroy, C., \& Speagle, J.~S. 2021,
  \bibinfo{title}{Stellar {Population} {Inference} with {Prospector},} The
  Astrophysical Journal Supplement Series, 254, 22,
  \dodoi{10.3847/1538-4365/abef67}

\bibitem[{A. {Kassiola} \& I. {Kovner}(1993){Kassiola} \&
  {Kovner}}]{KassiolaKovner1993}
{Kassiola}, A., \& {Kovner}, I. 1993, \bibinfo{title}{{Elliptic Mass
  Distributions versus Elliptic Potentials in Gravitational Lenses},} The
  Astrophysical Journal, 417, 450, \dodoi{10.1086/173325}

\bibitem[{L. Kawinwanichakij {et~al.}(2026)Kawinwanichakij, Glazebrook,
  Nanayakkara, Kacprzak, Chittenden, Jacobs, Chandro-G{\'o}mez, Lagos,
  Marchesini, Martìnez-Marìn, Oesch, \&
  Remus}]{kawinwanichakij_connecting_2026}
Kawinwanichakij, L., Glazebrook, K., Nanayakkara, T., {et~al.} 2026,
  \bibinfo{title}{Connecting {Environment}, {Star} {Formation} {History}, and
  {Morphology} of {Massive} {Quiescent} {Galaxies} at 3 {\textless} z
  {\textless} 4 with {JWST},} The Astrophysical Journal, 997, 29,
  \dodoi{10.3847/1538-4357/ae0a18}

\bibitem[{L.~J. Kewley {et~al.}(2013)Kewley, Dopita, Leitherer, Dav{\'e}, Yuan,
  Allen, Groves, \& Sutherland}]{kewley_theoretical_2013}
Kewley, L.~J., Dopita, M.~A., Leitherer, C., {et~al.} 2013,
  \bibinfo{title}{Theoretical {Evolution} of {Optical} {Strong} {Lines} across
  {Cosmic} {Time},} The Astrophysical Journal, 774, 100,
  \dodoi{10.1088/0004-637X/774/2/100}

\bibitem[{G. Khullar {et~al.}(2021)Khullar, Gozman, Lin, Martinez,
  Matthews~Acuña, Medina, Merz, Sanchez, Sisco, Kavin~Stein, Sukay, Tavangar,
  Bayliss, Bleem, Brownsberger, Dahle, Florian, Gladders, Mahler, Rigby,
  Sharon, \& Stark}]{khullarCOOLLAMPSExtraordinarilyBright2021}
Khullar, G., Gozman, K., Lin, J.~J., {et~al.} 2021,
  \bibinfo{title}{{COOL}-{LAMPS}. {I}. {An} {Extraordinarily} {Bright} {Lensed}
  {Galaxy} at {Redshift} 5.04,} The Astrophysical Journal, 906, 107,
  \dodoi{10.3847/1538-4357/abcb86}

\bibitem[{L.~C. Kimmig {et~al.}(2025)Kimmig, Remus, Seidel, Valenzuela, Dolag,
  \& Burkert}]{kimmig_blowing_2025}
Kimmig, L.~C., Remus, R.-S., Seidel, B., {et~al.} 2025, \bibinfo{title}{Blowing
  {Out} the {Candle}: {How} to {Quench} {Galaxies} at {High} {Redshift}---{An}
  {Ensemble} of {Rapid} {Starbursts}, {AGN} {Feedback}, and {Environment},} The
  Astrophysical Journal, 979, 15, \dodoi{10.3847/1538-4357/ad9472}

\bibitem[{M. Kriek \& C. Conroy(2013)Kriek \& Conroy}]{kriek_dust_2013}
Kriek, M., \& Conroy, C. 2013, \bibinfo{title}{The {Dust} {Attenuation} {Law}
  in {Distant} {Galaxies}: {Evidence} for {Variation} with {Spectral} {Type},}
  The Astrophysical Journal, 775, L16, \dodoi{10.1088/2041-8205/775/1/L16}

\bibitem[{C.~d.~P. Lagos {et~al.}(2024)Lagos, Bravo, Tobar, Obreschkow, Power,
  Robotham, Proctor, Hansen, Chandro-G{\'o}mez, \&
  Carrivick}]{lagos_quenching_2024}
Lagos, C. d.~P., Bravo, M., Tobar, R., {et~al.} 2024, \bibinfo{title}{Quenching
  massive galaxies across cosmic time with the semi-analytic model {SHARK}
  {V2}.0,} Monthly Notices of the Royal Astronomical Society, 531, 3551,
  \dodoi{10.1093/mnras/stae1024}

\bibitem[{S. Lebowitz {et~al.}(2025)Lebowitz, Hainline, Juneau, Lyu, Williams,
  Alberts, Fan, \& Rieke}]{lebowitz_jwst_2025}
Lebowitz, S., Hainline, K., Juneau, S., {et~al.} 2025, \bibinfo{title}{{JWST}
  {NIRCam} {Simulations} and {Observations} of {AGN} {Ionization} {Cones} in
  {Cosmic} {Noon} {Galaxies},} The Astrophysical Journal, 984, 13,
  \dodoi{10.3847/1538-4357/adc07c}

\bibitem[{B. Lorenz {et~al.}(2025)Lorenz, Suess, Kriek, Price, Leja, Nelson,
  Atek, Bezanson, Brammer, Cutler, Dayal, de~Graaff, Greene, Furtak, Labb{\'e},
  Marchesini, Maseda, Miller, Mintz, Mitsuhashi, Pan, Porraz~Barrera, Wang,
  Weaver, Williams, \& Whitaker}]{lorenzMeasuringEmissionLines2025}
Lorenz, B., Suess, K.~A., Kriek, M., {et~al.} 2025, \bibinfo{title}{Measuring
  {Emission} {Lines} with {JWST} {MegaScience} {Medium} {Bands}: {A} {New}
  {Window} into {Dust} and {Star} {Formation} at {Cosmic} {Noon},} The
  Astrophysical Journal Letters, 988, L20, \dodoi{10.3847/2041-8213/ade887}

\bibitem[{M. Mosleh {et~al.}(2017)Mosleh, Tacchella, Renzini, Carollo,
  Molaeinezhad, Onodera, Khosroshahi, \&
  Lilly}]{moslehConnectionStellarMass2017}
Mosleh, M., Tacchella, S., Renzini, A., {et~al.} 2017,
  \bibinfo{title}{Connection between {Stellar} {Mass} {Distributions} within
  {Galaxies} and {Quenching} {Since} z = 2,} The Astrophysical Journal, 837, 2,
  \dodoi{10.3847/1538-4357/aa5f14}

\bibitem[{L. {Mowla} {et~al.}(2024){Mowla}, {Iyer}, {Asada}, {Desprez}, {Tan},
  {Martis}, {Sarrouh}, {Strait}, {Abraham}, {Brada{\v{c}}}, {Brammer},
  {Muzzin}, {Pacifici}, {Ravindranath}, {Sawicki}, {Willott},
  {Estrada-Carpenter}, {Jahan}, {Noirot}, {Matharu}, {Rihtar{\v{s}}i{\v{c}}},
  \& {Zabl}}]{mowla_formation_2024}
{Mowla}, L., {Iyer}, K., {Asada}, Y., {et~al.} 2024, \bibinfo{title}{{Formation
  of a low-mass galaxy from star clusters in a 600-million-year-old Universe},}
  Nature, 636, 332, \dodoi{10.1038/s41586-024-08293-0}

\bibitem[{T. Naab {et~al.}(2009)Naab, Johansson, \& Ostriker}]{naab_minor_2009}
Naab, T., Johansson, P.~H., \& Ostriker, J.~P. 2009, \bibinfo{title}{Minor
  {Mergers} and the {Size} {Evolution} of {Elliptical} {Galaxies},} The
  Astrophysical Journal, 699, L178, \dodoi{10.1088/0004-637X/699/2/L178}

\bibitem[{T. Nanayakkara {et~al.}(2024)Nanayakkara, Glazebrook, Jacobs,
  Kawinwanichakij, Schreiber, Brammer, Esdaile, Kacprzak, Labbe, Lagos,
  Marchesini, Marsan, Oesch, Papovich, Remus, \&
  Tran}]{nanayakkara_population_2024}
Nanayakkara, T., Glazebrook, K., Jacobs, C., {et~al.} 2024, \bibinfo{title}{A
  population of faint, old, and massive quiescent galaxies at 3,} Scientific
  Reports, 14, 3724, \dodoi{10.1038/s41598-024-52585-4}

\bibitem[{A.~B. Newman {et~al.}(2015)Newman, Belli, \&
  Ellis}]{newman_discovery_2015}
Newman, A.~B., Belli, S., \& Ellis, R.~S. 2015, \bibinfo{title}{{DISCOVERY}
  {OF} {A} {STRONGLY} {LENSED} {MASSIVE} {QUIESCENT} {GALAXY} {AT} z = 2.636:
  {SPATIALLY} {RESOLVED} {SPECTROSCOPY} {AND} {INDICATIONS} {OF} {ROTATION},}
  The Astrophysical Journal Letters, 813, L7,
  \dodoi{10.1088/2041-8205/813/1/L7}

\bibitem[{A.~B. Newman {et~al.}(2018)Newman, Belli, Ellis, \&
  Patel}]{newmanResolvingQuiescentGalaxies2018a}
Newman, A.~B., Belli, S., Ellis, R.~S., \& Patel, S.~G. 2018,
  \bibinfo{title}{Resolving {Quiescent} {Galaxies} at z $\gtrsim$ 2. {II}.
  {Direct} {Measures} of {Rotational} {Support},} The Astrophysical Journal,
  862, 126, \dodoi{10.3847/1538-4357/aacd4f}

\bibitem[{Y. Ni {et~al.}(2025)Ni, Chen, Zhou, Park, Yang, Di~Matteo, Bird, \&
  Croft}]{ni_astrid_2025}
Ni, Y., Chen, N., Zhou, Y., {et~al.} 2025, \bibinfo{title}{The {Astrid}
  {Simulation}: {Evolution} of {Black} {Holes} and {Galaxies} to z = 0.5 and
  {Different} {Evolution} {Pathways} for {Galaxy} {Quenching},} The
  Astrophysical Journal, 990, 120, \dodoi{10.3847/1538-4357/adf3a7}

\bibitem[{J.~B. Oke \& J.~E. Gunn(1983)Oke \& Gunn}]{oke_secondary_1983}
Oke, J.~B., \& Gunn, J.~E. 1983, \bibinfo{title}{Secondary standard stars for
  absolute spectrophotometry,} The Astrophysical Journal, 266, 713,
  \dodoi{10.1086/160817}

\bibitem[{R.~G. Pascalau {et~al.}(2026)Pascalau, D'Eugenio, Tacchella,
  Maiolino, Cappellari, Duan, Lagos, Bunker, Jones, Scholtz, Übler, Cresci,
  Arribas, Perna, van~der Wel, Danhaive, McClymont, Williams, de~Graaff, Vani,
  Maseda, Carnall, Charlot, Carniani, Goh, Ji, \&
  P{\'e}rez~Gonz{\'a}lez}]{pascalauWhenRelicsWere2026}
Pascalau, R.~G., D'Eugenio, F., Tacchella, S., {et~al.} 2026,
  \bibinfo{title}{When relics were made: vigorous stellar rotation and low dark
  matter content in the massive ultra-compact galaxy {GS}-9209 at z = 4.66,}
  Monthly Notices of the Royal Astronomical Society, 547, stag210,
  \dodoi{10.1093/mnras/stag210}

\bibitem[{M. {Pascale} {et~al.}(2022){Pascale}, {Frye}, {Diego}, {Furtak},
  {Zitrin}, {Broadhurst}, {Conselice}, {Dai}, {Ferreira}, {Adams},
  {Kamieneski}, {Foo}, {Kelly}, {Chen}, {Lim}, {Meena}, {Wilkins},
  {Bhatawdekar}, \& {Windhorst}}]{pascale22}
{Pascale}, M., {Frye}, B.~L., {Diego}, J., {et~al.} 2022,
  \bibinfo{title}{{Unscrambling the Lensed Galaxies in JWST Images behind SMACS
  0723},} The Astrophysical Journal Letters, 938, L6,
  \dodoi{10.3847/2041-8213/ac9316}

\bibitem[{I. Pasha \& T.~B. Miller(2023)Pasha \& Miller}]{pasha_pysersic_2023}
Pasha, I., \& Miller, T.~B. 2023, \bibinfo{title}{pysersic: {A} {Python}
  package for determining galaxy structural properties via {Bayesian}
  inference, accelerated with jax,} The Journal of Open Source Software, 8,
  5703, \dodoi{10.21105/joss.05703}

\bibitem[{C.~Y. Peng {et~al.}(2002)Peng, Ho, Impey, \&
  Rix}]{peng_detailed_2002}
Peng, C.~Y., Ho, L.~C., Impey, C.~D., \& Rix, H.-W. 2002,
  \bibinfo{title}{Detailed {Structural} {Decomposition} of {Galaxy} {Images},}
  The Astronomical Journal, 124, 266, \dodoi{10.1086/340952}

\bibitem[{Y.-j. Peng {et~al.}(2010)Peng, Lilly, Kovač, Bolzonella, Pozzetti,
  Renzini, Zamorani, Ilbert, Knobel, Iovino, Maier, Cucciati, Tasca, Carollo,
  Silverman, Kampczyk, de~Ravel, Sanders, Scoville, Contini, Mainieri,
  Scodeggio, Kneib, Le~Fèvre, Bardelli, Bongiorno, Caputi, Coppa, de~la Torre,
  Franzetti, Garilli, Lamareille, Le~Borgne, Le~Brun, Mignoli, Perez~Montero,
  Pello, Ricciardelli, Tanaka, Tresse, Vergani, Welikala, Zucca, Oesch, Abbas,
  Barnes, Bordoloi, Bottini, Cappi, Cassata, Cimatti, Fumana, Hasinger,
  Koekemoer, Leauthaud, Maccagni, Marinoni, McCracken, Memeo, Meneux, Nair,
  Porciani, Presotto, \& Scaramella}]{pengMassEnvironmentDrivers2010}
Peng, Y.-j., Lilly, S.~J., Kovač, K., {et~al.} 2010, \bibinfo{title}{Mass and
  {Environment} as {Drivers} of {Galaxy} {Evolution} in {SDSS} and {zCOSMOS}
  and the {Origin} of the {Schechter} {Function},} The Astrophysical Journal,
  721, 193, \dodoi{10.1088/0004-637X/721/1/193}

\bibitem[{R.-S. Remus \& L.~C. Kimmig(2025)Remus \&
  Kimmig}]{remusRelightCandleWhat2025}
Remus, R.-S., \& Kimmig, L.~C. 2025, \bibinfo{title}{Relight the {Candle}:
  {What} {Happens} to {High}-redshift {Massive} {Quenched} {Galaxies},} The
  Astrophysical Journal, 982, 30, \dodoi{10.3847/1538-4357/ad8b4b}

\bibitem[{M. Revalski {et~al.}(2025)Revalski, Crenshaw, Polack, Rafelski,
  Kraemer, Fischer, Meena, Schmitt, Trindade~Falcão, Falcone, \&
  Shea}]{revalski_quantifying_2025}
Revalski, M., Crenshaw, D.~M., Polack, G.~E., {et~al.} 2025,
  \bibinfo{title}{Quantifying {Feedback} from {Narrow} {Line} {Region}
  {Outflows} in {Nearby} {Active} {Galaxies}. {V}. {The} {Expanded} {Sample},}
  The Astrophysical Journal, 984, 32, \dodoi{10.3847/1538-4357/adc131}

\bibitem[{N. Roy {et~al.}(2024)Roy, Heckman, Overzier, Saxena, Duncan, Miley,
  Villar~Martín, Gab{\'a}nyi, Aydar, Bosman, Rottgering, Pentericci, Onoue, \&
  Reynaldi}]{roy_jwst_2024}
Roy, N., Heckman, T., Overzier, R., {et~al.} 2024, \bibinfo{title}{{JWST}
  {Reveals} {Powerful} {Feedback} from {Radio} {Jets} in a {Massive} {Galaxy}
  at z = 4.1,} The Astrophysical Journal, 970, 69,
  \dodoi{10.3847/1538-4357/ad4bda}

\bibitem[{B. Salmon {et~al.}(2020)Salmon, Coe, Bradley, Bouwens, Bradač,
  Huang, Oesch, Stark, Sharon, Trenti, Avila, Ogaz, Andrade-Santos, Carrasco,
  Cerny, Dawson, Frye, Hoag, Johnson, Jones, Lam, Lovisari, Mainali, Past,
  Paterno-Mahler, Peterson, Riess, Rodney, Ryan, Sendra-Server, Strait,
  Strolger, Umetsu, Vulcani, \& Zitrin}]{salmon_relics_2020}
Salmon, B., Coe, D., Bradley, L., {et~al.} 2020, \bibinfo{title}{{RELICS}:
  {The} {Reionization} {Lensing} {Cluster} {Survey} and the {Brightest}
  {High}-z {Galaxies},} The Astrophysical Journal, 889, 189,
  \dodoi{10.3847/1538-4357/ab5a8b}

\bibitem[{P. S{\'a}nchez-Bl{\'a}zquez {et~al.}(2006)S{\'a}nchez-Bl{\'a}zquez,
  Peletier, Jim{\'e}nez-Vicente, Cardiel, Cenarro, Falcón-Barroso, Gorgas,
  Selam, \& Vazdekis}]{sanchez-blazquez_medium-resolution_2006}
S{\'a}nchez-Bl{\'a}zquez, P., Peletier, R.~F., Jim{\'e}nez-Vicente, J.,
  {et~al.} 2006, \bibinfo{title}{Medium-resolution {Isaac} {Newton} {Telescope}
  library of empirical spectra,} Monthly Notices of the Royal Astronomical
  Society, 371, 703, \dodoi{10.1111/j.1365-2966.2006.10699.x}

\bibitem[{N. Saputra~Haryana {et~al.}(2026)Saputra~Haryana, Akiyama,
  {Abdurro'uf}, Cooray, Khoirul~Fitriana, Triani, Vijarnwannaluk, Alfonzo,
  Taufik~Andika, Effendi, Nurul~Huda, Kakimoto, Timur~Jaelani, Laishram,
  Matsumoto, Naufal, Puspitarini, Sutanto, \&
  Wulandari}]{saputraharyanaInterplayCompactionQuenching2026}
Saputra~Haryana, N., Akiyama, M., {Abdurro'uf}, {et~al.} 2026, Interplay of
  {Compaction}, {Quenching}, and {Black} {Hole} {Growth} in the {Most}
  {Massive} {Galaxies} since $z\sim5$: {Insights} from {JWST} and {Chandra}
  {Data}, arXiv, \dodoi{10.48550/arXiv.2608.24124}

\bibitem[{A. Saxena {et~al.}(2024)Saxena, Overzier, Villar-Martín, Heckman,
  Roy, Duncan, R{\"o}ttgering, Miley, Aydar, Best, Bosman, Cameron,
  Gab{\'a}nyi, Humphrey, Morais, Onoue, Pentericci, Reynaldi, \&
  Venemans}]{saxena_widespread_2024}
Saxena, A., Overzier, R.~A., Villar-Martín, M., {et~al.} 2024,
  \bibinfo{title}{Widespread {AGN} feedback in a forming brightest cluster
  galaxy at z = 4.1, unveiled by {JWST},} Monthly Notices of the Royal
  Astronomical Society, 531, 4391, \dodoi{10.1093/mnras/stae1406}

\bibitem[{D.~J. Setton {et~al.}(2024)Setton, Khullar, Miller, Bezanson, Greene,
  Suess, Whitaker, Antwi-Danso, Atek, Brammer, Cutler, Dayal, Feldmann,
  Fujimoto, Furtak, Glazebrook, Goulding, Kokorev, Labbe, Leja, Ma, Marchesini,
  Nanayakkara, Pan, Price, Siegel, Shipley, Weaver, van Dokkum, Wang, \&
  Williams}]{settonUNCOVERNIRSpecPRISM2024}
Setton, D.~J., Khullar, G., Miller, T.~B., {et~al.} 2024,
  \bibinfo{title}{{UNCOVER} {NIRSpec}/{PRISM} {Spectroscopy} {Unveils}
  {Evidence} of {Early} {Core} {Formation} in a {Massive}, {Centrally} {Dusty}
  {Quiescent} {Galaxy} at z spec = 3.97,} The Astrophysical Journal, 974, 145,
  \dodoi{10.3847/1538-4357/ad6a18}

\bibitem[{J.~C. Siegel {et~al.}(2025)Siegel, Setton, Greene, Suess, Whitaker,
  Bezanson, Leja, Furtak, Cutler, de~Graaff, Feldmann, Khullar, Labbe,
  Marchesini, Miller, Nanayakkara, Pan, Price, Treiber, van Dokkum, Wang, \&
  Weaver}]{siegel_uncover_2025}
Siegel, J.~C., Setton, D.~J., Greene, J.~E., {et~al.} 2025,
  \bibinfo{title}{{UNCOVER}: {Significant} {Reddening} in {Cosmic} {Noon}
  {Quiescent} {Galaxies},} The Astrophysical Journal, 985, 125,
  \dodoi{10.3847/1538-4357/adc7b7}

\bibitem[{R.~S. Somerville \& R. Davé(2015)Somerville \&
  Davé}]{somervillePhysicalModelsGalaxy2015}
Somerville, R.~S., \& Davé, R. 2015, \bibinfo{title}{Physical {Models} of
  {Galaxy} {Formation} in a {Cosmological} {Framework},} Annual Review of
  Astronomy and Astrophysics, 53, 51,
  \dodoi{10.1146/annurev-astro-082812-140951}

\bibitem[{J.~S. Speagle(2020)Speagle}]{speagle_dynesty_2020}
Speagle, J.~S. 2020, \bibinfo{title}{{DYNESTY}: a dynamic nested sampling
  package for estimating {Bayesian} posteriors and evidences,} Monthly Notices
  of the Royal Astronomical Society, 493, 3132, \dodoi{10.1093/mnras/staa278}

\bibitem[{K.~A. Suess {et~al.}(2020)Suess, Kriek, Price, \&
  Barro}]{suess_color_2020}
Suess, K.~A., Kriek, M., Price, S.~H., \& Barro, G. 2020, \bibinfo{title}{Color
  {Gradients} along the {Quiescent} {Galaxy} {Sequence}: {Clues} to {Quenching}
  and {Structural} {Growth},} The Astrophysical Journal, 899, L26,
  \dodoi{10.3847/2041-8213/abacc9}

\bibitem[{P.~G. van Dokkum {et~al.}(2010)van Dokkum, Whitaker, Brammer, Franx,
  Kriek, Labbé, Marchesini, Quadri, Bezanson, Illingworth, Muzzin, Rudnick,
  Tal, \& Wake}]{vandokkumGrowthMassiveGalaxies2010}
van Dokkum, P.~G., Whitaker, K.~E., Brammer, G., {et~al.} 2010,
  \bibinfo{title}{The {Growth} of {Massive} {Galaxies} {Since} z = 2,} The
  Astrophysical Journal, 709, 1018, \dodoi{10.1088/0004-637X/709/2/1018}

\bibitem[{P.~G. van Dokkum {et~al.}(2015)van Dokkum, Nelson, Franx, Oesch,
  Momcheva, Brammer, F{\"o}rster~Schreiber, Skelton, Whitaker, van~der Wel,
  Bezanson, Fumagalli, Illingworth, Kriek, Leja, \&
  Wuyts}]{vandokkumFormingCompactMassive2015}
van Dokkum, P.~G., Nelson, E.~J., Franx, M., {et~al.} 2015,
  \bibinfo{title}{Forming {Compact} {Massive} {Galaxies},} The Astrophysical
  Journal, 813, 23, \dodoi{10.1088/0004-637X/813/1/23}

\bibitem[{E. {Vanzella} {et~al.}(2022){Vanzella}, {Castellano}, {Bergamini},
  {Meneghetti}, {Zanella}, {Calura}, {Caminha}, {Rosati}, {Cupani},
  {Me{\v{s}}tri{\'c}}, {Brammer}, {Tozzi}, {Mercurio}, {Grillo}, {Sani},
  {Cristiani}, {Nonino}, {Merlin}, \& {Pignataro}}]{vanzella_high_2022}
{Vanzella}, E., {Castellano}, M., {Bergamini}, P., {et~al.} 2022,
  \bibinfo{title}{{High star cluster formation efficiency in the strongly
  lensed Sunburst Lyman-continuum galaxy at z = 2.37},} Astronomy and
  Astrophysics, 659, A2, \dodoi{10.1051/0004-6361/202141590}

\bibitem[{E. {Vanzella} {et~al.}(2023){Vanzella}, {Claeyssens}, {Welch},
  {Adamo}, {Coe}, {Diego}, {Mahler}, {Khullar}, {Kokorev}, {Oguri},
  {Ravindranath}, {Furtak}, {Hsiao}, {Abdurro'uf}, {Mandelker}, {Brammer},
  {Bradley}, {Brada{\v{c}}}, {Conselice}, {Dayal}, {Nonino}, {Andrade-Santos},
  {Windhorst}, {Pirzkal}, {Sharon}, {de Mink}, {Fujimoto}, {Zitrin},
  {Eldridge}, \& {Norman}}]{vanzella_jwstnircam_2023}
{Vanzella}, E., {Claeyssens}, A., {Welch}, B., {et~al.} 2023,
  \bibinfo{title}{{JWST/NIRCam Probes Young Star Clusters in the Reionization
  Era Sunrise Arc},} The Astrophysical Journal, 945, 53,
  \dodoi{10.3847/1538-4357/acb59a}

\bibitem[{J.~R. Weaver {et~al.}(2023)Weaver, Davidzon, Toft, Ilbert, McCracken,
  Gould, Jespersen, Steinhardt, Lagos, Capak, Casey, Chartab, Faisst, Hayward,
  Kartaltepe, Kauffmann, Koekemoer, Kokorev, Laigle, Liu, Long, Magdis,
  McPartland, Milvang-Jensen, Mobasher, Moneti, Peng, Sanders, Shuntov,
  Sneppen, Valentino, Zalesky, \& Zamorani}]{weaverCOSMOS2020GalaxyStellar2023}
Weaver, J.~R., Davidzon, I., Toft, S., {et~al.} 2023,
  \bibinfo{title}{{COSMOS2020}: {The} galaxy stellar mass function. {The}
  assembly and star formation cessation of galaxies at 0.2{\textless} z $\leq$
  7.5,} Astronomy and Astrophysics, 677, A184,
  \dodoi{10.1051/0004-6361/202245581}

\bibitem[{J.~R. Weaver {et~al.}(2024)Weaver, Cutler, Pan, Whitaker, Labb{\'e},
  Price, Bezanson, Brammer, Marchesini, Leja, Wang, Furtak, Zitrin, Atek,
  Chemerynska, Coe, Dayal, van Dokkum, Feldmann, F{\"o}rster~Schreiber, Franx,
  Fujimoto, Fudamoto, Glazebrook, de~Graaff, Greene, Juneau, Kassin, Kriek,
  Khullar, Maseda, Mowla, Muzzin, Nanayakkara, Nelson, Oesch, Pacifici,
  Papovich, Setton, Shapley, Shipley, Smit, Stefanon, Taylor, Weibel, \&
  Williams}]{weaver_uncover_2024}
Weaver, J.~R., Cutler, S.~E., Pan, R., {et~al.} 2024, \bibinfo{title}{The
  {UNCOVER} {Survey}: {A} {First}-look {HST} + {JWST} {Catalog} of 60,000
  {Galaxies} near {A2744} and beyond,} The Astrophysical Journal Supplement
  Series, 270, 7, \dodoi{10.3847/1538-4365/ad07e0}

\bibitem[{A. Weibel {et~al.}(2025)Weibel, de~Graaff, Setton, Miller, Oesch,
  Brammer, Lagos, Whitaker, Williams, Baggen, Bezanson, Boogaard, Cleri,
  Greene, Hirschmann, Hviding, Kuruvanthodi, Labb{\'e}, Leja, Maseda, Matthee,
  McConachie, Naidu, Roberts-Borsani, Schaerer, Suess, Valentino, van Dokkum,
  \& Wang}]{weibel_rubies_2025}
Weibel, A., de~Graaff, A., Setton, D.~J., {et~al.} 2025,
  \bibinfo{title}{{RUBIES} {Reveals} a {Massive} {Quiescent} {Galaxy} at z =
  7.3,} The Astrophysical Journal, 983, 11, \dodoi{10.3847/1538-4357/adab7a}

\bibitem[{K.~E. Whitaker \& R. Bezanson(2026)Whitaker \&
  Bezanson}]{whitaker_quenching_2026}
Whitaker, K.~E., \& Bezanson, R. 2026, \bibinfo{title}{Quenching of {Star}
  {Formation} in {Massive} {Galaxies},} Annual Review of Astronomy and
  Astrophysics, 64, 179, \dodoi{10.1146/annurev-astro-043024-121502}

\bibitem[{K.~E. Whitaker {et~al.}(2017)Whitaker, Bezanson, van Dokkum, Franx,
  van~der Wel, Brammer, F{\"o}rster-Schreiber, Giavalisco, Labb{\'e}, Momcheva,
  Nelson, \& Skelton}]{whitaker_predicting_2017}
Whitaker, K.~E., Bezanson, R., van Dokkum, P.~G., {et~al.} 2017,
  \bibinfo{title}{Predicting {Quiescence}: {The} {Dependence} of {Specific}
  {Star} {Formation} {Rate} on {Galaxy} {Size} and {Central} {Density} at 0.5
  {\textless} z {\textless} 2.5,} The Astrophysical Journal, 838, 19,
  \dodoi{10.3847/1538-4357/aa6258}

\bibitem[{A. {Zitrin} {et~al.}(2015){Zitrin}, {Fabris}, {Merten}, {Melchior},
  {Meneghetti}, {Koekemoer}, {Coe}, {Maturi}, {Bartelmann}, {Postman},
  {Umetsu}, {Seidel}, {Sendra}, {Broadhurst}, {Balestra}, {Biviano}, {Grillo},
  {Mercurio}, {Nonino}, {Rosati}, {Bradley}, {Carrasco}, {Donahue}, {Ford},
  {Frye}, \& {Moustakas}}]{Zitrin2015}
{Zitrin}, A., {Fabris}, A., {Merten}, J., {et~al.} 2015,
  \bibinfo{title}{{Hubble Space Telescope Combined Strong and Weak Lensing
  Analysis of the CLASH Sample: Mass and Magnification Models and Systematic
  Uncertainties},} The Astrophysical Journal, 801, 44,
  \dodoi{10.1088/0004-637X/801/1/44}

\bibitem[{A. Zolotov {et~al.}(2015)Zolotov, Dekel, Mandelker, Tweed, Inoue,
  DeGraf, Ceverino, Primack, Barro, \&
  Faber}]{zolotovCompactionQuenchingHighz2015}
Zolotov, A., Dekel, A., Mandelker, N., {et~al.} 2015,
  \bibinfo{title}{Compaction and quenching of high-z galaxies in cosmological
  simulations: blue and red nuggets,} Monthly Notices of the Royal Astronomical
  Society, 450, 2327, \dodoi{10.1093/mnras/stv740}

\end{thebibliography}

\end{document}